\documentclass[11pt]{article}
\usepackage{graphicx}
\usepackage{amsmath,amsfonts,amssymb}
\usepackage{color}
\usepackage{bbold}
\usepackage{braket}
\allowdisplaybreaks
\def\sloppy{\tolerance=100000\hfuzz=\maxdimen\vfuzz=\maxdimen}
\allowdisplaybreaks
\def \beq  {\begin{equation}}
\def \eeq  {\end{equation}}
\def \beqar {\begin{eqnarray}}
\def \eeqar {\end{eqnarray}}
\def\sqr#1#2{{\vcenter{\vbox{\hrule height.#2pt
\hbox{\vrule width.#2pt height#1pt \kern#1pt
\vrule width.#2pt}\hrule height.#2pt}}}}

\mathchardef\mhyphen="2D
\DeclareMathOperator\arctanh{arctanh}

\def\L {{\cal L}}

\def\la {{\langle}}
\def\ra {{\rangle}}

\def\vf {{\varphi}}

\def\Tr {{\rm Tr}}

\def\bcalE {\bar{\cal E}}

\def\bk{{\bar k}}

\def\bn{{\bar n}}

\def\bs {\bar{s}}

\def\bw {\bar{w}}

\def\vf {{\varphi}}

\def\bxi {{\bar \xi}}

\def\del {\partial}
\def\bdel{\bar{\partial}}

\def\D {{\cal D}}

\def\A {{\cal A}}

\def\E {{\cal E}}

\def\H {{\cal H}}
\def \L {{\cal L}}
\def\M{{\cal M}}

\def\W {{\cal W}}
\def\bcalW {{\overline{\cal W}}}

\def\half{\textstyle{1\over 2}}

\begin{document}
%%%%%%%%%%%%%%%%%%%%%%%%%%%%%%%%%%%%%%%%%%%%%%%%
%%%%%%%%%%%%%%%%%%%%%%%%%%%%%%%%%%%%%%%%%%%%%%%%
%%%%%%%%%%%%%%%%%%%%%%%%%%%%%%%%%%%%%%%%%%%%%%%%
\fontfamily{bch}\fontsize{11pt}{14.6pt}\selectfont
%\fontfamily{pnb}\fontsize{12pt}{16pt}\selectfont
%\fontfamily{pzc}\fontsize{14pt}{16pt}\selectfont
%\fontfamily{pbk}\fontsize{12pt}{16pt}\selectfont
%\fontfamily{cmr}\fontsize{11pt}{15pt}\selectfont
%\fontfamily{phv}\fontshape{ro}\fontsize{11pt}{14pt}\selectfont
%\fontfamily{ptm}\fontseries{m}\fontshape{r}\fontsize{12pt}{16pt}\selectfont
%\fontfamily{pnc}\fontseries{m}\fontshape{r}\fontsize{11pt}{15pt}\selectfont
%\fontfamily{ppl}\fontseries{m}\fontshape{r}\fontsize{11pt}{15pt}\selectfont
%\usefont{T1}{phv}{m}{it}
%%%%%%%%%%%%%%%%%%%%%%%%%%%%%%%%%%%%%%%%%%%%%%%
\def \CMP {{Commun. Math. Phys.}}
\def \PRL {{Phys. Rev. Lett.}}
\def \PL {{Phys. Lett.}}
\def \NPBProc {{Nucl. Phys. B (Proc. Suppl.)}}
\def \NP {{Nucl. Phys.}}
\def \RMP {{Rev. Mod. Phys.}}
\def \JGP {{J. Geom. Phys.}}
\def \CQG {{Class. Quant. Grav.}}
\def \MPL {{Mod. Phys. Lett.}}
\def \IJMP {{ Int. J. Mod. Phys.}}
\def \JHEP {{JHEP}}
\def \PR {{Phys. Rev.}}
\def \JMP {{J. Math. Phys.}}
\def \GRG{{Gen. Rel. Grav.}}
%%%%%%%%%%%%%%%%%%%%%%%%%%%%%%%%%%%%%%%%%%%%%%%
%%%%%%%%%%%%%%%%%%%%%%%%%%%%%%%%%%%%%%%%%%%%%%%
%%%%%%%%%%%%%%%%%%%%%%%%%%%%%%%%%%%%%%%%%%%%%%%
%%%%%%%%%%%%%%%%%%%%%%%%%%%%%%%%%%%%%%%%%%%%%%%
\begin{titlepage}
\null\vspace{-62pt} \pagestyle{empty}
\begin{center}
%\rightline{CCNY-HEP-18-02}
%\rightline{February 2018}
\vspace{1truein} {\Large\bfseries
A Note on Coherent States for Virasoro Orbits}\\
\vspace{6pt}
\vskip .1in
{\Large \bfseries  ~}\\
%\vskip .2in
{\Large\bfseries ~}\\
%%%%%%%%%%%%%%%%%%%%%%%%%%%%%%%%%%%%%%%%%%%%%%%%\vspace{.6in}
 {\large\sc V.P. Nair}\\
\vskip .2in
{\itshape Physics Department\\
City College of the City University of New York\\
New York, NY 10031}\\
\vskip .1in
{\itshape School of Theoretical Physics\\
Dublin Institute for Advanced Studies\\
10 Burlington Road, Dublin 4, Ireland}\\
\vskip .1in
\begin{tabular}{r l}
E-mail:&\!\!\!{\fontfamily{cmtt}\fontsize{11pt}{15pt}\selectfont vpnair@ccny.cuny.edu}
\end{tabular}

\vspace{.8in}
%\vspace{1.5in}%\vspace{0.3in}
\centerline{\large\bf Abstract}
\end{center}
There are two sets of orbits of the Virasoro group which admit a K\"ahler structure. We consider
the construction of coherent states for the
orbit {$\widehat{{\rm diff}\,S^1}$/$SL(2, {\mathbb R})$} which furnishes unitary representations of the
group. The procedure is analogous to geometric quantization
using a holomorphic polarization.
We also give an explicit formula for the K\"ahler potential
for this orbit and comment on normalization of the
coherent states.
We further explore some of the properties of these states, including the definition of symbols
corresponding to operators and their star products.
Some comments which touch upon the
possibility of applying this to gravity in (2+1) dimensions are also given.

\end{titlepage}

%%%%%%%%%%%%%%%%%%%%%%%%%%%%%%%%%%%%%%%%%%%%%%%%
%%%%%%%%%%%%%%%%%%%%%%%%%%%%%%%%%%%%%%%%%%%%%%%%
%%%%%%%%%%%%%%%%%%%%%%%%%%%%%%%%%%%%%%%%%%%%%%%%
%%%%%%%%%%%%%%%%%%%%%%%%%%%%%%%%%%%%%%%%%%%%%%%%
\pagestyle{plain} \setcounter{page}{2}

\section{Introduction}
It has been well known for a long time that one can construct coherent states 
which realize unitary representations of a Lie group via the method
of geometric
quantization \cite{reviews}. For this one considers coadjoint orbits of the Lie group
$G$ which are of the form $G/H$ for a suitable subgroup $H$ such that
the coset space has a K\"ahler structure.
For the general case one chooses $H$ to be the maximal torus
in $G$, with the K\"ahler two-form given by
$-\sum_1^r w_i \Tr ( h_i g^{-1} dg \wedge g^{-1} dg )$
where $g$ denotes a general group element in the fundamental
representation (viewed as a matrix), $h_i$ are the generators of
the maximal torus in a suitable basis and $(w_1, w_2, \cdots, w_r)$ defines
the highest weight state of some unitary representation of $G$,
and $r$ denotes the rank of the group.
The result of the quantization will be a Hilbert space $\H$ corresponding to the
carrier space of the representation with the highest weight indicated.
One can assign wave functions to the states of this Hilbert space; they obey
a holomorphicity condition depending on the K\"ahler potential.
Orthonormality conditions, symbols corresponding to operators
on $\H$, star products, the diagonal coherent state representation, etc. can then be defined
in a straightforward way.

In this article, we consider a similar construction of coherent states, star products, etc. for
the Virasoro group, which will be identified as the centrally extended version of 
diffeomorphisms of a circle, denoted 
{{$\widehat{{\rm diff}\,S^1}$}}.
This problem is clearly of some intrinsic mathematical interest, but there are several motivating factors from physics as well. After all, the Virasoro algebra is one of the foundational ingredients for
the formulation of string theory. Partly motivated by this, coadjoint orbits of 
{{$\widehat{{\rm diff}\,S^1}$}} were classified and some of their properties analyzed
many years ago \cite{{witten1},{bowick, rajeev}}.
Another context in which {{$\widehat{{\rm diff}\,S^1}$}} emerges is 
(2+1)-dimensional gravity.
The action for this theory (with a cosmological constant)
is given as the difference of two $SL(2, {\mathbb R})$
Chern-Simons actions, with the connection forms
$A_L$, $A_R$ given as combinations of the frame fields and spin connection.
Witten's analysis of the partition function of this theory shows that the inclusion of the
BTZ black holes will require Virasoro representations in the relevant sum over states
\cite{{witten2}, {maloney}}.
The construction of the coherent states is useful for such analyses.
Further, the semiclassical limit of this analysis corresponds to taking the central charge
$c$ to be large, a limit which is suitable for a star-product expansion for observables.

In the case of finite dimensional Lie groups, the Hilbert space $\H$ can be used
as a model for a description of the noncommutative version of the manifold
$G/H$. Coherent states are then useful in defining symbols corresponding to
operators on $\H$ and the star products give the (noncommutative) algebra of functions
on $G/H$. Given the appearance of Virasoro representations
in (2+1)-dimensional gravity, if
we envisage a noncommutative antecedent for gravity,
then coherent states for the Virasoro algebra become important in defining symbols and 
star products and obtaining a continuous manifold description in the large
$c$ limit.

In the case of (2+1)-dimensional gravity on asymptotically
anti-de Sitter space, the asymptotic symmetries also
lead to a Virasoro algebra
\cite{brown-henneaux}.
Some of the issues of gravity may thus be cast in terms of
holography or the AdS/CFT correspondence, which is a different facet of string theory. 
Combined with the
observations in the previous paragraphs, this suggests a way to bring together ideas of 
string theory and/or gravity and noncommutative geometry.
The coherent states for the Virasoro group
will also be central to any such attempt.

Among the coadjoint orbits of the Virasoro group, there are two which admit
K\"ahler structures and hence are amenable to defining coherent states
satisfying appropriate holomorphicity conditions. These orbits correspond to
{{$\widehat{{\rm diff}\,S^1}$/$S^1$}} and 
{$\widehat{{\rm diff}\,S^1}$/$SL(2, {\mathbb R})$}. The K\"ahler potentials
for these cases are characterized by two numbers, the central charge
$c$, which must be viewed as the eigenvalue of the
central element of the algebra, and $h$, which may be viewed as the
eigenvalue of the generator $L_0$ on the highest weight state
of the representation.
For most purposes, the quantization of
{$\widehat{{\rm diff}\,S^1}$/$SL(2, {\mathbb R})$} may be considered as a
special case of the quantization of 
{{$\widehat{{\rm diff}\,S^1}$/$S^1$}}, so we will focus on the latter space.
Some of the other orbits are relevant for representations which
contain null vectors and can lead to unitary representations
for $c < 1$. We will not discuss these here.

Not surprisingly, there have been many discussions in the literature
which touch upon various aspects of coherent states for the Virasoro group.
Many of the papers we have cited contain some implicit statements
about such states.
Representations of the Virasoro algebra as differential operators on
functions of a suitable set of variables occurs
in the context of the KP hierarchy and the so-called string
equation \cite{DVV}. A different, but related, representation has been used
in discussions of the stochastic Loewner equation \cite{bauer}.
These do not directly lead to coherent states.
The group manifold approach to quantization, discussed in
\cite{aldaya}, is a generalization of the idea of the
coadjoint orbit quantization and as such is closer to our discussion.
This approach starts with the formal group law and imposes polarization conditions
on functions on the group to obtain irreducible representations.
In \cite{aldaya}, the composition laws for the Virasoro group parameters,
including the central parameter relevant to the central extension, have
been constructed. 
Our approach is more traditional, but clearly there are some points of 
overlap; we will briefly comment on this later.
For us 
holomorphicity is important as it provides a simple way to construct
 a suitable
star product. (The Kontsevich formula for star products does not need
coherent states {\it per se}, but it does need more information about the symplectic
structure on the space \cite{kontsevich}.
 In turn, this would require an analysis similar to
what we do in this article, so it does not seem like our
analysis can be evaded.)
Standard coherent states for the bosonic
operators obeying the Heisenberg algebra in the mode expansion for
the target space coordinate in string theory  have also been considered 
in the literature \cite{skliros}. While these can be useful for
certain applications, the states we are concerned with are not these;
we are interested in directly quantizing the Virasoro orbits.

This paper is organized as follows. In the next section we will briefly discuss coherent
states for $SL(2, {\mathbb R})/U(1)$. The results here are well-known, but 
it helps to set the stage for a similar analysis for the Virasoro case, which is taken up in section 3.
The key steps are the following. 
We first construct an expression for the K\"ahler potential
for the orbit of interest. This will give a functional version of the
coherent states and it will also inform the issues of normalization
discussed later.
We then define
an operator $U$ and show that
it is possible to choose coordinates on the Virasoro orbit
such that $U^{-1} dU$ can be split into holomorphic and antiholomorphic forms corresponding to the Virasoro generator $L_{-n}$ and its conjugate.
This naturally leads to a set of wave functions associated with
the states of the Verma module and which obey a certain holomorphicity condition.
The normalization of these wave functions will involve the
K\"ahler potential. 
The normalization integral with the appropriate measure
of integration is then discussed.
Symbols and star products are considered in section 4.
In the limit of large central charge, the star products reduce to those
on $SL(2, \mathbb{R})/U(1)$, i.e., on a noncommutative version of
${\rm AdS}_2$.
The paper concludes with a short
discussion in section 5. There is also
an Appendix which gives details of some of the assertions made in
section 3.

%%%%%%%%%%%%%%%%%%%%%%%%%%%%%%%%%%%%%%%%%%%%%%%
%%%%%%%%%%%%%%%%%%%%%%%%%%%%%%%%%%%%%%%%%%%%%%%
\section{Coherent states for $SL(2, {\mathbb R})/U(1)$}

In this section, we will briefly consider the construction of coherent states
for $SL(2, {\mathbb R})$. While these results are well known, our presentation will
help to highlight certain points which can help
to clarify the lines of argument employed later for
 the Virasoro case.

The generators of the Lie algebra can be taken as $L_0$, $L_{\pm 1}$ with the 
commutation rules
\beq
[L_0, L_{\pm 1} ] = \mp \, L_{\pm 1}, \hskip .3in
[L_1, L_{-1} ] = 2\, L_0 .
\label{sl1}
\eeq
The highest weight state is defined by
\beq
L_1 \, \ket{0} = 0, \hskip .3in L_0 \, \ket{0} =  h\, \ket{0} .
\label{sl2}
\eeq
We will consider representations with $h > {\half}$ which are the simplest 
for exemplifying
the arguments used for the Virasoro case.
Other states are obtained by the action of powers of $L_{-1}$ on $\ket{0}$.
The normalized states are given by
\beq
\ket{n} = {1\over \sqrt{N(n)}} L_{-1}^n \ket{0},
\hskip .2in
N(n) = {\Gamma (n+1) \Gamma( 2 h + n) \over \Gamma( 2 h )},
\label{sl3}
\eeq
where $\Gamma (u) $ is Eulerian gamma function for the argument $u$.
 The state $\ket{n}$ corresponds
to a value of $L_0$ equal to $h + n$, so the representation we
are considering is bounded below.
Now we introduce the unitary operator\footnote{The coordinates
$w$, $\bw$ correspond to what we later call $w_1$, $\bw_1$
in the context of the Virasoro algebra; we drop the subscripts for this section to avoid
clutter.}
\beq
U = \exp\left( w \, L_{-1} - {\bar w} \, L_1 \right),
\hskip .2in U^\dagger = \exp\left({\bar w} \, L_1  -  w \, L_{-1} \right) ,
\label{sl4}
\eeq
where $w$,${\bar w}$ are functions of some complex coordinates $s$, $\bs$ which parametrize
$SL(2, {\mathbb R})/U(1)$.
We now define the coherent states by the wave functions
\beq
\Psi_n = \bra{0} U^\dagger \ket{n} = {1\over \sqrt{N(n)}} \bra{0} U^\dagger \, L_{-1}^n \ket{0} .
\label{sl4a}
\eeq
(A similar definition will be used later for the Virasoro case.)

For the case of $SL(2, {\mathbb R})$, a
general element of the group can be written in the
$2\times 2$ matrix representation as
\beq
g = {1\over \sqrt{1- \bs s}} \left( \begin{matrix} 1& i s\\ -i \bs& 1\\ \end{matrix} \right)
\, \left( \begin{matrix} e^{i\vf /2}&0\\ 0&e^{-i \vf/2}\\ \end{matrix} \right), \hskip .2in 
\vert s\vert < 1 .
\label{sl5}
\eeq
The correspondence of the Lie algebra elements in this representation
 is given by
$L_0 = {\half} \sigma_3$, $L_1 = {i\over 2}( \sigma_1 - i \sigma_2)$,
$L_{-1} =  {i\over 2} (\sigma_1 +i \sigma_2 )$, in terms of the standard Pauli matrices.
The group parameters
$s$, $\bs$ provide local coordinates (for a patch around $s = 0$) for the coset space
$SL(2, {\mathbb R})/ U(1)$. For this parametrization,
the standard K\"ahler
one-form $\Tr (\sigma_3 g^{-1} dg )$ will contain a term $d\vf$, which will drop out in
the two-form $d (\Tr (\sigma_3 g^{-1} dg ))$. So, for the quantization of the orbit
corresponding to the chosen $h$-value on $SL(2, {\mathbb R}/U(1)$, we can drop $\vf $
on the local coordinate patch.
The parametrization also shows a singularity at $\vert s\vert = 1$; this is integrable for
$h > \half$, and will not affect results below. 

For the parametrization in (\ref{sl5}), for a unitary representation
of $g$ as $U$, we get
\beqar
d U^\dagger \, U^{\dagger -1} &=& \left( {s d\bs - \bs ds \over 1- \bs s}\, L_0 
+ {d \bs \over 1- \bs s} \, L_1 - { ds \over 1- \bs s} \, L_{-1} \right)\nonumber\\
U^{\dagger -1} d U^\dagger &=&  \left( - {s d\bs - \bs ds \over 1- \bs s}\, L_0 
+ {d \bs \over 1- \bs s} \, L_1 - { ds \over 1- \bs s} \, L_{-1} \right) .
\label{sl6}
\eeqar
From the first of these relations, and using (\ref{sl2}), we get the holomorphicity conditions for the
coherent states as
\beq
\left( {\del \over \del s} + { h\, \bs \over 1- \bs s} \right) \Psi_n = 0 .
\label{sl7}
\eeq
The second equation in (\ref{sl6}) also gives
\beq
\left( {\del \over \del \bs} + { h\, s \over 1- \bs s} \right) \Psi_0 = 0 .
\label{sl9}
\eeq
For $\Psi_0$ we can solve these equations to find
$ \bra{0} U^\dagger \ket{0} = (1- \bs s)^h$. The possible arbitrary multiplicative
constant is set to one, since $U^\dagger  = 1$ at $s = 0$.
For the higher states, we can develop a recursion rule using (\ref{sl7}) as follows.
\beqar
\Psi_n &=&  {1\over \sqrt{N(n)}} \bra{0} U^\dagger \, L_{-1}\, L_{-1}^{n-1} \ket{0}\nonumber\\
&=&{1\over \sqrt{N(n)}} \bra{0} \left( - (1- \bs s) {\del U^\dagger \over \del s} + U^\dagger \, \bs\, L_0 \right)
L_{-1}^{n-1} \, \ket{0}\nonumber\\
&=&\sqrt{N(n-1)\over N(n)}  \left( - (1- \bs s) {\del  \over \del s} + \, \bs\,(h + n -1) \right) \Psi_{n-1} .
\label{sl10}
\eeqar
This can be solved to write
\beq
\Psi_n =  \sqrt{\Gamma(2 h +n ) \over \Gamma (n+1) \Gamma(2 h)}
~ \bs^n \, (1- \bs s)^h .
\label{sl11}
\eeq

The K\"ahler potential may be identified from the symplectic form
or from the holomorphicity condition written as $(\del_s + {\half} \del_s K ) \Psi_n = 0$.
 We can also identify $K$ from
the relation  $\bra{0} U^{-1} dU \ket{0} = {\half} (\del K - \bdel K)$.
The K\"ahler potential and the symplectic
form can be worked out as
\beq
K = - 2 \, h \log (1- \bs s), \hskip .2in
\omega =  i \del \bdel K = 2\, h  \, i{d\bs \, ds \over (1- \bs s)^2} \, .
\label{sl12}
\eeq
We will use a slightly different normalization 
 for the phase volume defined by $\omega$ (with a prefactor $2h-1$ rather than $2h$)
 which makes
the $\Psi_n$ orthonormal. It is easy to verify that
\beq
(2 h -1)  \int {d^2s \over \pi (1-\bs s)^2 }
~\Psi_n^* \, \Psi_m = \delta_{nm} .
\label{sl13}
\eeq
The range of integration is over the disk $\vert s\vert \leq 1$;
the singularity at $\vert s\vert = 1$ is integrable for $h > {\half}$, so the
inner product is well-defined.

In equations (\ref{sl7}, \ref{sl11}, \ref{sl13}), we
have reproduced the standard and well-known results for coherent states
on $SL(2, {\mathbb R})/ U(1)$. 
For $SL(2, {\mathbb R})$ we have the advantage of a parametrization for the
coset space given as in (\ref{sl5}); for the Virasoro case, we have to rely
on a power series expansion for $w$, $\bw$.
We will determine the expressions for $w$, $\bw$ as a series in $s$, $\bs$
relying on the K\"ahler property and requiring that
the coefficient of $L_{-1}$ be a holomorphic one-form.
We will briefly illustrate the strategy here.
Notice that, using (\ref{sl4}), we can write
\beqar
U^\dagger d U &=& \int_0^1 d\alpha ~
e^{- \alpha (w L_{-1} - \bw L_1 )} \, (dw L_{-1} - d\bw L_1 ) \,
e^{\alpha (w L_{-1} - \bw L_1 )} \nonumber\\
&\approx& \left[ dw \left(1 + {1\over 3} \bw w \right)  - d\bw\, {1\over 3} w^2  \right]\, L_{-1}
- \left[d\bw \left(1 + {1\over 3} \bw w \right) - dw \,{1\over 3} \bw^2 \right]
\, L_{1} \nonumber\\
&&+ (\bw dw - w d\bw ) \, L_0 + \cdots .
\label{sl14}
\eeqar
We take $w$ to be given in terms of the local coordinates as
\beq
w = s + w^{(2)}(s, \bs) + w^{(3)}(s, \bs) +\cdots .
\label{sl15}
\eeq
Setting the coefficient of $d\bs \, L_{-1}$ to zero we find
$w^{(2)} = s^2 \bs /3$. Using this, we can simplify $dU^\dagger \, U = - U^\dagger dU$
 as
\beq
 dU^\dagger \,U =   d\bs \, (1+ \bs s) \, L_1 - ds\, (1+ \bs s) \, L_{-1} 
+ (s d\bs - \bs ds )\, L_0 + \cdots .
\label{sl16}
\eeq
This matches with the expression in (\ref{sl6}) to first order in the expansion
in powers of $\bs s$.
We have checked that the process can be continued by developing $w$ as a series
in $s$, $\bs$ and requiring that the coefficient of $L_{-1}$ is a holomorphic one-form,
to reproduce the results in (\ref{sl6}). 
We will use a similar strategy for
the Virasoro group, developing a series expansion for $w_n$, $\bw_n$, as worked out in the Appendix.
Since the procedure is exactly parallel, the
results we obtain in that case will revert to the discussion in the present section when 
restricted to the coordinates $w_1$, $\bw_1$, and to $L_0, \, L_{\pm 1}$, setting
$w_n$, $\bw_n = 0$, $n \geq 2$.

In the case of $SL(2, {\mathbb R})$, one can also directly use the Baker-Campbell-Hausdorff
(BCH) formula
\beqar
e^{ w L_{-1} - \bw L_1} &=&
e^{s L_{-1}}\, 
e^{ \log (1 - \bs s) L_0} \, e^{-\bs L_1}\nonumber\\
s&=& {w \over \vert w\vert} \tanh \vert w\vert
 \label{sl17}
\eeqar
Since the BCH formula is determined by the commutator algebra of 
$L_0$, $L_{\pm 1}$, the matrix representation $L_0 = \sigma_3 /2$,
$L_{\pm 1} = i (\sigma_1 \pm i \sigma_2)/2$ can be used to verify
(\ref{sl17}). Writing $w$ in terms of $s$ we get
$w = (s/ \vert s\vert) \arctanh \vert s\vert \approx s + s^2\bs/3 + \cdots$,
in agreement with the series expansion in (\ref{sl15}).
Further by taking the expectation value of the first equation in (\ref{sl17}) for the highest weight state
$\ket{0}$, we find
\beq
\la 0 \vert U^\dagger  \vert 0\ra = 
\la 0\vert e^{- s L_{-1}}\, 
e^{ \log (1 - \bs s) L_0} \, e^{- \bs L_1}\vert 0\ra = 
e^{h \log (1- \bs s)} \equiv e^{-K/2}
\label{sl18}
\eeq
which agrees with the expression for the K\"ahler potential in (\ref{sl12}).

Finally, we also note that $K \rightarrow \infty$ at the boundary of the region of
integration, which is $\vert s\vert \rightarrow 1$. This corresponds to
$\vert w\vert \rightarrow \infty$ if we use the coordinates $w,\, \bw$.

\section{Construction of the states for the Virasoro case}

As mentioned in the Introduction, unitary representations of the Virasoro group, for {$c > 1$}, are obtained
by quantizing {{$\widehat{{\rm diff}\,S^1}$/$S^1$}} and 
{$\widehat{{\rm diff}\,S^1}$/$SL(2, {\mathbb R})$}. Further, these orbits have a 
K\"ahler structure \cite{{witten1},{bowick, rajeev}}.
We will focus on  {{$\widehat{{\rm diff}\,S^1}$/$S^1$}} for reasons mentioned earlier.
 This manifold can be described by complex coordinates
 $s_k$ and ${\bar s}_{k}$, where $k$  takes integer values from
 $1$ to infinity.
 The basic generators of the Virasoro algebra are $L_n$, $n \in {\mathbb Z}$,
with the commutator algebra
\beq
[ L_n, L_m ] = (n -m )\, L_{n+m} + {c \over 12} (n^3 - n) \,\delta_{n+m, 0}\, {\mathbb 1} .
\label{1}
\eeq
Here ${\mathbb 1}$ is the identity operator; the central charge
may be viewed as the eigenvalue
of a central charge operator ${\hat c}$ which is proportional to the identity.
This is necessary to view the algebra (\ref{1}) as having a closed Lie algebra
structure. Thus there is a slight abuse of notation in writing
(\ref{1}) directly in terms of the eigenvalue; this will be immaterial
for what we want to do.
 We are interested in highest weight representations, with
 the highest weight state
 obeying
 \beq	
 L_0 \, \ket{0} = h\, \ket{0}, \hskip .2in
 L_n \, \ket{0} = 0, \hskip .2in n \geq 1 .
 \label{2}
\eeq
Further, we will be interested in the case of
$c > 1$ and $h > 0$, so that we will not have null vectors in the
associated Verma module.
From now on we will use the notation $L_0$, $L_n$, $L_{-n}$, with
$n \geq 1$ explicitly distinguishing the three sets of operators.
The subgroup $S^1$ in {{$\widehat{{\rm diff}\,S^1}$/$S^1$}} is generated by
the action of $L_0$, with $L_n$ , $L_{-n}$ forming the translation
operators on the coset {{$\widehat{{\rm diff}\,S^1}$/$S^1$}}.

\subsection{The K\"ahler ptential}

Some considerations on the nature of the K\"ahler potential
will be useful before embarking on constructing explicit formulae for the coherent states.
 An expression for the symplectic two-form relevant for
 {$\widehat{{\rm diff}\,S^1}$/$SL(2, {\mathbb R})$}
 was given by Stanford and Witten in \cite{S-W} as
 \beq
 \omega = {1\over 4\pi} \int_0^{2\pi}  d\tau\,
 \left[ {c \over 12} {\delta \phi' \over \phi'} {\del \over \del \tau}
 \left( {\delta \phi' \over \phi'}\right)  - {c\over 12}
 \delta \phi \,\delta \phi'  \right]
 \label{KP1}
 \eeq
 Here $\phi (\tau)$ may be considered as
 a field on the circle parametrized by
 $0\leq \tau \leq 2\pi$. The prime denotes differentiation with respect to 
 $\tau$, Explicitly, $\phi$ is of the form
 \beqar
 \phi (\tau) &=& \tau + \vf + {\bar \vf} +\chi + {\bar \chi} \nonumber\\
\vf&=&  = s_1 e^{i \tau}, \hskip .2in
{\bar \vf} = {\bar s}_1 e^{-i \tau}\nonumber\\
 \chi &=&  \sum_2^\infty s_n e^{i n \tau},  \hskip .2in
{\bar \chi} =  \sum_2^\infty
 {\bar s}_n e^{-i n \tau} 
 \label{KP2}
 \eeqar
Further $\delta$ in (\ref{KP1}) denotes exterior derivative on the space of the fields $\phi$ and wedge products for the $\delta$'s is understood.
The two-form $\omega$ has zero modes corresponding to
$L_0$, $L_\pm 1$, i.e, for $n = 0,  \pm1$.
This is in accordance with the fact that $\omega$
is the symplectic form
for {$\widehat{{\rm diff}\,S^1}$/$SL(2, {\mathbb R})$}.
The zero modes, as argued in \cite{S-W}, correspond to the vector fields
\beq
V_n = \int d\tau\, e^{i n \phi(\tau )} {\delta \over \delta \phi (\tau)}, 
\hskip .2in n = 0, \pm 1
\label{KP2a}
\eeq
The zero modes can be removed by a suitable gauge-fixing condition
to obtain an $\omega$ which is invertible.
In \cite{S-W}, this was chosen as $\phi (0) = 0$, $\phi'(0) = 1$ and
$\phi''(0) = 0$. For us, it will be convenient to make a different choice.
Under the action of $V_n$, with parameters $\beta_0, \beta_{\pm1}$,
the field $\phi$ transforms as
\beqar
\phi &\rightarrow& \phi + \beta_0 + \beta_{+1} e^{i \phi} + \beta_{-1} e^{-i \phi}
\nonumber\\
&=&\tau + \vf +{\bar \vf} + \beta_0 + \beta_{+1} e^{i \tau} e^{i (\vf + {\bar \vf})} +  \beta_{-1} e^{- i \tau} e^{-i (\vf + {\bar \vf})}\nonumber\\
&\approx& \tau + \vf +{\bar \vf} + \beta_0 + \beta_{+1} e^{i \tau}  +  \beta_{-1} e^{- i \tau }
+\cdots
\label{KP2b}
\eeqar 
where, in the last line, we have expanded out $e^{ \pm i (\vf + {\bar \vf})}$.
This shows that the coordinates $s_1$, $\bs_1$ shift by $\beta_{+1}$,
$\beta_{-1}$ for a neighborhood around $s_n = 0$. Even though the full transformation will be nonlinear, this shows that a suitable gauge-fixing for
$V_{\pm 1}$ is to set $s_1 = \bs_1 = 0$.
The gauge-fixing for $V_0$ is taken care of by setting
$\phi = \tau$ for $s_n =0$. Thus a gauge-fixed version of
$\omega$ is given by the same expression as in (\ref{KP1}),
but with 
\beq
\phi = \tau + \chi + {\bar \chi},
 \label{KP2c}
 \eeq
omitting the $s_1$, $\bs_1$ terms.

The gauge-fixing provides a local section of ${\rm diff}\,S^1
\sim \bigl[ {{\rm diff}\,S^1}$/$SL(2, {\mathbb R})\bigr]\times SL(2, \mathbb{R})$.
For  {$\widehat{{\rm diff}\,S^1}/U(1)$}, we have to modify $\omega$
 to take care of the term corresponding to
$SL(2, \mathbb{R})/ (U(1)$. A suitable modification is given by
 \beq
 \omega = {1\over 4\pi} \int_0^{2\pi}  d\tau\,
 \left[ {c \over 12} {\delta \phi' \over \phi'} {\del \over \del \tau}
 \left( {\delta \phi' \over \phi'}\right) +\left(2 h - {c\over 12}\right)
 \delta \phi \,\delta \phi'  - 4 h {\delta{\bar\vf}' \, \delta \vf \over (1- i {\bar\vf}' \vf )^2}\right]
 \label{KP2d}
 \eeq
 We have made two changes compared to (\ref{KP1}). 
 We have added the contribution due to $s_1$, ${\bar s}_1$
 which is the part due to $SL(2, \mathbb{R})/ (U(1)$.
Secondly, recall that the $n^3$-term in the central term of
the algebra (\ref{1}) is the cohomologically nontrivial term.
This corresponds to the first term in $\omega$.
The term linear in $n$, which corresponds to the
integral of $\delta \phi \delta \phi'$, can be modified by changing the value of
$L_0$.
Accordingly, to take account of the $h$-dependence for the
modes $s_n$, $\bs_n$, $n \geq 2$, we have added a term
proportional to $2 h \delta \phi \delta \phi'$.
 
If we substitute from (\ref{KP2c}) into (\ref{KP2d}), $\omega$ will 
have terms of the form
$\delta \chi' \, \delta\chi''$ and its conjugate, i.e., of the
$(2,0)$ and $(0,2)$ types of differential forms.
We can separate out the terms which correspond to the $(1,1)$-type
 as
 \beqar
 \omega&=& \omega_{\rm K} + F\nonumber\\
 \omega_{\rm K} &=& {1\over 4\pi} \int_0^{2\pi}  d\tau\,
\left\{{c \over 12} \left[  {\delta {\bar\chi}' \, \delta \chi'' -\delta {\bar\chi}'' \, \delta \chi' \over  (1+ \chi' +{\bar \chi}')^2}  + 2 {({\bar \chi}'' - \chi'') \delta{\bar\chi}' \, \delta \chi' \over (1+ \chi' +{\bar \chi}')^3}\right]\right.
\nonumber\\
&&\hskip .9in \left.+ \left(4 h - {c\over 6} \right)\, \delta{\bar\chi} \, \delta \chi'
 - 4 h {\delta {\bar \vf}' \, \delta \vf \over
 (1- i {\bar\vf}'\vf)^2} \right\}\label{KP3}\\
 F&=& {1\over 4\pi} \int_0^{2\pi}  d\tau\,
\left\{{c \over 12} \left[  {\delta \chi'\, \delta\chi'' + \delta {\bar\chi}' \, \delta {\bar\chi}'' \over ( 1+ \chi' + {\bar \chi}')^2} - 2 {({\bar \chi}'' - \chi'') \delta{\bar\chi}' \, \delta \chi' \over (1+ \chi' +{\bar \chi}')^3}\right]\right\}
\label{KP4}
 \eeqar
 It is also easy to see that we can write $F = d A = (\delta + {\bar\delta}) A$, where
 \beq
 A = -{1\over 4\pi} {c \over 12} \int_0^{2\pi}  d\tau\,
{\chi'' \, \delta \chi' + {\bar\chi}'' \, \delta {\bar \chi}' \over (1+ \chi' +{\bar\chi}')^2}
\label{KP5}
\eeq
The two-form $F$ in (\ref{KP3}) can be removed by a coordinate transformation.
This is essentially the idea of Moser's lemma. Even though we have an infinite number of modes in $\chi$, $\bar\chi$, Moser's argument should apply since $\omega$ is invertible. We define $\omega_t =
\omega_{\rm K} + t \,F$. The condition that $\omega_{t+\epsilon}
-\omega_t$ is given by an infinitesimal coordinate
transformation by the vector field $X_t$ becomes
\beq
{d \omega_t \over dt} = \L_{X_t} \omega_t =
d (i_{X_t} \omega_t )
\label{KP5a}
\eeq
where $\L$ denotes the Lie derivative.
This equation reduces to
\beq
d ( A - i_{X_t} \omega_t ) = 0
\label{KP5b}
\eeq
In terms of the components corresponding to
$s_n$, $\bs_n$, the solution is given by
\beq
X_t^k = (\omega_t)^{k l} A_l
\label{KP5c}
\eeq
There are no zero modes for $\omega_t$, so the inverse should exist, albeit it is an infinite dimensional matrix.
This shows how the required coordinate transformation can be constructed.
Effectively, we can connect $\omega_{\rm K} $ (which is $ \omega_t $ at $t = 0$)
to $\omega$ (which is $ \omega_t $ at $t =1$) by a series of coordinate transformations.
(A series expansion around
$ s_n, {\bar s}_n = 0$ may be needed for
constructing the explicit formulae for the
required coordinate transformation.) The key point is that $\omega$ 
in (\ref{KP3}) is thus symplectomorphic
to $\omega_{\rm K}$ which is of the $(1,1)$-type.
We can therefore choose $\omega_{\rm K}$ as the symplectic form for our
problem.

It is now easy to work out the K\"ahler potential for $\omega_{\rm K}$.
It is given by
\begin{align}
K = {1\over 4\pi} \int_0^{2\pi}  \!\!\!d\tau\,
& \Biggl\{  {c \over 12} \left[
 -i({{\bar\chi}''- \chi'' )\over (1+ \chi' +{\bar\chi}')}\right] + \left( 4 h - {c\over 6}\right)  (-i  {\bar\chi}\, \chi' )
  - 4 h \log(1- i {\bar\vf}' \vf)
\Biggr\}\nonumber\\
= {1\over 4\pi} \int_0^{2\pi}  \!\!\!d\tau\,
& \Biggl\{ {c \over 12} \left[
 {i \over (1+ \chi' +{\bar\chi}')}\left( { {\bar\chi}'' \chi' \over (1+ {\bar\chi}')}
- {\chi'' {\bar \chi}' \over (1+ \chi')} \right)\right]\nonumber\\
& + \left( 4 h - {c\over 6}\right) (- i \,{\bar \chi}\, \chi' )
 - 4 h \log(1- i {\bar\vf}' \vf) \Biggr\}
 \label{KP6}
\end{align}
In passing to the second line of this equation, we
used the fact that the integral of expressions which are purely holomorphic
in the fields, i.e., depending only on $\chi$ 
(or antiholomorphic involving only  ${\bar\chi}$) vanish upon $\tau$-integration.
Using (\ref{KP6}), it is straightforward to check that $\omega_{\rm K} = i {\bar\delta} \delta K$.
The expansion of this expression for small $s_n, \bs_n$ will also agree with
the alternate derivation of $K$, as will be clear from the discussion in
subsection 3.2, equation (\ref{9a}). 
Also, if we restrict to just the modes $s_1$, $\bs_1$, corresponding to
$SL(2, \mathbb{R})$, it coincides with the expression
for $K$ in (\ref{sl12}). If we restrict to $s_n$, ${\bar s}_n$ for fixed
choice of $n$, corresponding to the subgroup defined by
$L_0$, $L_{\pm n}$, the resulting expression is not of the same form as
for $s_1$, $\bs_1$. A further coordinate transformation will be needed to
bring it to the form of the K"ahler potential for $SL(2, \mathbb{R})/ U(1)$
with $s_1, \bs_1 \rightarrow s_n, \bs_n$.

Once we have the explicit form of the K\"ahler potential, one can construct coherent states as is usually done for any K\"ahler manifold.
In the functional form, these are given by
\beq
\Psi [\bar\vf] = {\cal N} \, e^{- {\half} K} \, \Phi [{\bar\vf}, {\bar \chi} ]
\label{KP7}
\eeq
We will now construct another expression for the coherent states
in terms of the states of the Verma module, the final result
being (\ref{16d}). We expect that (\ref{KP7}) will coincide
with (\ref{16d}), up to a possible coordinate transformation, or, equivalently, a redefinition of fields, once it is projected to the
states of the Verma module.

\subsection{The operator $U$ and the choice of complex coordinates}

While (\ref{KP7}) is formally a definition of the coherent states,
a more explicit formula in terms of the states of the 
Verma module, similar to the construction of the
$SL(2, \mathbb{R})/U(1)$ states in section 2, will also be useful.
As a first step towards this, we define a unitary operator $U$ by
\beq
U = \exp \left( \sum_{n =1}^\infty {\bar w}_n  \, L_n - w_n L_{-n} \right) .
 \label{3}
\eeq
We will be considering unitary highest weight
representations, so $U$ will be unitary.
In (\ref{3}), $w_n$ and $\bw_n$ are functions of the coordinates $s_k, {\bs}_k$.
Generally, from the commutation rules, we can see that we can write
\beq
U^{-1} dU =  \sum_n   (\bcalE^n\, L_n - {\cal E}^n L_{-n} ) +
({\cal E}^0 - {\bar{\cal E}}^0 ) \, L_0 + ( {\cal E} - {\bar{\cal E}} ) {\mathbb 1} ,
 \label{4}
\eeq
where the coefficients $\E^n$, $\E^0$, $\E$ and their conjugates are one-forms
on the coset space, with components which are functions
of the coordinates. In other words, we can write
\beq
\E^n = \sum_k ( \E^n_k \, ds_k + \E^n_{\bk} d \bs_k), 
\hskip .2in
\bcalE^n = \sum_k (\bcalE^n_k \, ds_k + \bcalE^n_{\bk} d \bs_k) .
 \label{5}
\eeq
As for the coefficients of $L_0$ and $\mathbb 1$, we take
 $\E^0$, $\E$ to be holomorphic one-forms, with
$\bcalE^0$, $\bcalE$ being the conjugates.

Our construction of the coherent states will follow the steps we outlined for
the case of $SL(2, \mathbb{R})$. We will first show 
that $\E^n$ can be chosen to be a holomorphic one-form
and $\bcalE^n$ as an antiholomorphic one-form.
We can then define coherent states which will obey an antiholomorphicity condition. We will then set up and
analyze the normalization of the wave functions for the coherent states.
This will give us the ingredients for the symbols and star products taken up in the next section.

The first key result we will need is that it is possible to choose 
$w_n$, $\bw_n$ as functions of the coordinates in such a way that
$\sum_k \E^n_\bk d\bs_k = 0$ and $\sum_k \bcalE^n_k ds_k = 0$.\footnote{
An issue with notation: To distinguish the holomorphic and antiholomorphic
pieces of the
one-forms $\E^n$, $\bcalE^n$, we use subscripts with an overbar. Once we argue
that $\E^n_\bk $ and $\bcalE^n_k$ are zero, we will drop this distinction 
to avoid the notational clutter of writing $\bs^{\bk}$, etc.}
In other words,
$\E^n$ is a holomorphic one-form and $\bcalE^n$ is an antiholomorphic one-form,
\beq
\E^n = \sum_k  \E^n_k\, ds_k  ,  \hskip .2in \bcalE^n = \sum_k \bcalE^n_k \, d\bs_k , \hskip .2in
\E^n_{\bar k} = \bcalE^n_k = 0.
 \label{6}
\eeq
This result can be established by expansion around the origin and then using
homogeneity of the orbit. 
The key point is that we can consider $w_n$ and $\bw_n$ to be defined by a 
power series expansion
in terms of the coordinates $s_k$, $\bs_k$, the coefficients of the expansion
can be fixed by requiring $\E^n_k d\bs_k = 0$ and $\bcalE^n_k ds_k = 0$.
To the quadratic order in the coordinates, we find
\beq
\bw_n = \bs_{n} - {1\over 2} \sum_m (n + 2m) \bs_{{n} + {m}} s_m
+ \cdots ,
 \label{7}
\eeq
with $w_n$ given by the complex conjugate.
(The details of the required calculations are given in the Appendix.)
Thus for a small neighborhood around the origin in the chosen coordinates
we can see that we do obtain the holomorphicity conditions (\ref{6}).
To this order, we then find
\beq
({\cal E}^0 - {\bar{\cal E}}^0 ) \, L_0 + ( {\cal E} - {\bar{\cal E}} ) {\mathbb 1} =
- {\half} \sum_n ( s_n d{\bar s}_{ n} - {\bar s}_n ds_n )\left[
2 n L_0 + {c\over 12} (n^3 - n) \right] + \cdots  .
 \label{8}
\eeq
If we evaluate this on the highest weight state, we find
\beqar
\left( ({\cal E}^0 - {\bar{\cal E}}^0 ) \, L_0 + ( {\cal E} - {\bar{\cal E}} ) {\mathbb 1} 
\right) \, \ket{0}&=&{1\over 2} \sum_k \left( {\del \W \over \del s_k} ds_k -{\del \bcalW \over \del \bs_k} d\bs_k \right) \ket{0}
 \nonumber\\
&=& {1\over 2} (\del \W - \bdel \bcalW )\, \ket{0}
\label{9}
\eeqar
Here $\del$ and $\bdel$ are the $(1,0)$ and $(0,1)$ components of the
exterior derivative.
We show in the Appendix that $\E^0 = {\half} \del W^0$, $\E = {\half} \del W$ 
for some functions $W^0$ and $W$ and
$\W = W^0 h + W$ in (\ref{9}). As we will see from the K\"ahler two-form given later, the K\"ahler potenial is given by
$K = {\half} ( \W + \bcalW )$. To the order we have evaluated these functions in
(\ref{8}),
\beq
K =  {1\over 2} (\W + {\bar \W} ) \approx \sum_n  s_n \bs_n \,\left[ 2 n h + {c\over 12} (n^3 - n) \right] + \cdots  .
 \label{9a}
\eeq
The fact that the K\"ahler potential
$K$ is characterized by two numbers
$h$ and $c$ has been observed and commented on before
\cite{{witten1}, {bowick, rajeev}}. In the Appendix, we give the expansion of
$\E^n$ and $\bcalE^n$ to the next order in the coordinates.

To go beyond the small neighborhood around the origin, we will use the homogeneity of
the coset space. Assume that we have obtained the result (\ref{6}) in some neighborhood of the
origin. We then
consider translations of 
$U$ as $U \, V$ where $V$ is of the form
\beq
V = \exp \left( \sum_{n =1}^\infty {\bar \xi}_n  \, L_n - \xi_n L_{-n} \right) ,
 \label{10}
\eeq
for infinitesimal $\xi_n$, ${\bar \xi}_n$.  Writing out $(U V)^{-1} d (U V)$
to first order in $\xi_n$, ${\bar \xi}_n$, 
the holomorphicity conditions (\ref{6}) become differential
equations for these quantities. The integrability conditions
for these equations are satisfied by virtue of the
Maurer-Cartan identities for
$U^{-1} dU$.
Therefore, one
can extend the neighborhood where
the holomorphicity conditions are obtained.
The detailed calculations supporting these statements are given in the Appendix.

\subsection{Coherent states}

We will now move to the next step in
the construction of the coherent states.
Starting from the highest weight state $\ket{0}$, we can define
the states in the Verma module of the form
\beq
\ket{\{ {\widetilde n} \} } =
\cdots  L_{-3}^{n_3} L_{-2}^{n_2} L_{-1}^{n_1} \, \ket{0} ,
 \label{11}
\eeq
where we can use the Virasoro algebra to order the
$L_{-n}$'s in increasing level number to the left.
(We use $\{ {\widetilde n}\}$ with a tilde over $n$ to denote the unnormalized states;
for the normalized states, given below, we will write
$\{ n \}$.)
The matrix of inner products for the (unnormalized) states of the Verma module is given
as
\beq
\M_{ \{n\}, \{m \} } =  \braket{\{\widetilde n \} |\{ \widetilde m \} } = \bra{0} L_1^{n_1} L_2^{n_2} L_3^{n_3} \cdots
  L_{-3}^{m_3} L_{-2}^{m_2} L_{-1}^{m_1} \, \ket{0} .
   \label{12}
\eeq
This has a block-diagonal form, with the inner product between states of different
level number being zero.
The properly normalized states are then
\beq
\ket{\{ n \} } = \sum_{\{m\}} (\M^{-\half} )_{\{n \} , \{m \} }\,
\cdots  L_{-3}^{m_3} L_{-2}^{m_2} L_{-1}^{m_1} \, \ket{0} .
 \label{13}
\eeq
We can now define the coherent state wave function corresponding to the state
$\ket{\{ n \} } $ by
\beq
\Psi_{\{ n\} } =  \la 0 \vert U^\dagger | \{ n \} \ra \, e^{i \Theta}
 \label{14}
\eeq
where $\Theta$ is a phase to be specified shortly.
Using $U^\dagger dU = - dU^\dagger U$ and  (\ref{4}), write\footnote{We expect
that the
one-form $U^\dagger dU$ is closely related to a similar quantity
in reference 9, which develops and uses
the group manifold approach to quantization of the Virasoro group.
In this approach, the group law is first obtained recursively starting from the commutation
rules. Left and right group actions can then be
obtained on functions on the group manifold. The resulting representations are reducible
in general. Subsidiary conditions, the polarization conditions, are chosen to obtain
irreducible representations. The approach is more general than, but closely related to, the standard
geometric quantization.  From the group law, the so-called quantization one-form
is constructed, equation (3.4) of that paper. Holomorphicity is not {\it a priori} important, since
one is using the group law. However, the two cases of ``complete polarization" considered in
\cite{aldaya} do reduce to the two orbits with K\"ahler structure, 
{{$\widehat{{\rm diff}\,S^1}$}}/$U(1)$ and {{$\widehat{{\rm diff}\,S^1}$}}/$SL(2, {\mathbb R})$.
We expect that there is a transformation of the coordinates $l_k$ used
in \cite{aldaya} to our $s_k, \bs_k$, by which the one-form
in (3.4) of that paper reduces to our result (\ref{15}), up to a total differential.
For us holomorphicity is important in developing the star product, so we have
chosen to separate the coordinates into holomorphic and anti-holomorphic ones
from the beginning so that we have holomorphicity for the one-forms
$\E$.}
\beq
 d U^\dagger =  \left[ \sum_{n,k}  - \bcalE^n_{k} d\bs_k  \, L_n +  \E^n_k ds_k\, L_{-n}
 - \left( (\E^0 - \bcalE^0) L_0 - (\E - \bcalE) {\mathbb 1} \right)\right]\, U^\dagger  .
  \label{15}
\eeq
This shows that the wave functions (\ref{14}) obey the 
 antiholomorphicity condition
 \beq
 \left( {\del \over \del s_k} + {1\over 2} {\del \W \over \del s_k} +i {\del \Theta \over \del s_k} \right)
 \Psi_{\{ n \} } = 0 .
\label{16}
\eeq
We now choose $\Theta = i (\W - {\bar \W} )/2$, so that
this equation becomes
 \beq
 \left( {\del \over \del s_k} + {1\over 2} {\del K \over \del s_k}  \right)
 \Psi_{\{ n \} } = 0 .
\label{16f}
\eeq
where $K = {\half}(\W + {\bar \W})$ is the K\"ahler potential.
This equation can be solved for the coherent state wave functions
to write
\beq
\Psi_{\{n\} } = \la 0 \vert U^\dagger | \{ n \} \ra\, e^{i \Theta} 
= e^{-{\half} K } \, \Phi_{\{n\} } (\bs )
\label{16d}
\eeq
where $\Phi_{\{n\} }(\bs )$ depend only on the antiholomorphic coordinates
$\bs_k$.

\subsection{The K\"ahler two-form and the phase volume}

Our next step is to write down the K\"ahler two-form and the
K\"ahler potential in terms of $U$, for the purpose of setting up the normalization integrals for the states (\ref{16d}).

Going back to (\ref{4}), notice that it defines a  left-invariant one-form $U^{-1} dU$,
since $(V_L U )^{-1} d (V_L U) = U^{-1} dU$ for constant $V_L$. Further,
\beq
\bra{0} U^{-1} dU \ket{0} = {1\over 2} (\del \W - \bdel\, \bcalW ) .
\label{17}
\eeq
This leads to a left-invariant two-form, which is the K\"ahler two-form
and which can serve as the symplectic structure
of interest, given by
\beq
\omega = i\, d \bra{0} U^{-1} dU \ket{0} =  {i \over 2} \bdel \del (\W + \bcalW ) =
i\,\bdel \del K .
\label{16a}
\eeq
We also note
that the existence of a left-invariant symplectic structure has been emphasized by
Witten \cite{witten1}. Since $d (U^{-1} dU ) = - U^{1-} dU\, U^{-1} dU$, we can use
(\ref{15}) to obtain another useful expression for $\omega$,
\beqar
\omega &=& - i\bra{0}  U^{1-} dU\, U^{-1} dU \ket{0}
\nonumber\\
&=&- i \sum_{n,k} \bra{0} \Bigl[ \left( \bcalE^n L_n - \E^n L_{-n}  + 
(\E^0 - \bcalE^0) L_0 + (\E - \bcalE) {\mathbb 1}\right)\nonumber\\
&&\hskip .6in \wedge 
\left( \bcalE^k L_k - \E^k L_{-k}  + 
(\E^0 - \bcalE^0) L_0 + (\E - \bcalE) {\mathbb 1}\right) \Bigr] \ket{0}
\nonumber\\
&=& i \sum_{n,k} \bcalE^n\wedge  \E^k\, \bra{0} L_n L_{-k} \ket{0}
\nonumber\\
&=&i \sum_n \bcalE^n\wedge  \E^n\, \left( 2 n h + {c \over 12} (n^3 - n) \right) 
\equiv i \, \Omega_{kl}\, d\bs_k \wedge ds_l
\label{16b}\\
\Omega_{kl} &=& \sum_n \bcalE^n_{k}   \E^n_l \, \left( 2 n h + {c \over 12} (n^3 - n) \right)
\nonumber
\eeqar
The K\"ahler one-form $\A$ corresponding to this is given by
(\ref{17}) as 
\beq
\A =  {i \over 2} (\del \W - \bdel\, \bcalW ) 
= {i \over 2} (\del K - \bdel K ) + d (\Theta/ 2)
\label{16c}
\eeq
The antiholomorphicity condition (\ref{16f}) may be viewed as the polarization condition for geometric quantization of $\omega$ in (\ref{16a}) or (\ref{16b}).
The exact term in $\A$ is removed by a phase transformation of the
wave functions,. as we have already done in defining 
$\Psi_n$ in (\ref{14}).

The K\"ahler two-form defines
a phase-space  volume $d\mu$ or an integration measure which is
invariant under constant left translations of $U$. This may be written as
\beq
d\mu = (\det \Omega ) \, \prod_k d\bs_k \, ds_k
\label{16e}
\eeq
Since there are an infinite number of coordinates,
any integration carried out using this
must be understood in a regularized sense,
as defined over a finite set of modes,
taking the limit of the total number of modes becoming infinite at the end.
The determinant of $\Omega$ must be understood in a similar regularized 
way.
\subsection{Normalization of wave functions}

With the  understanding of the left-invariant measure
as given above, we can now show that the coherent states (\ref{14}) can be 
normalized. We will first show the reasoning for the orthonormality
of the wave functions, assuming the existence of the relevant
integrals. The latter issue will be taken up subsequently.
We start by considering the normalization integral
\beqar
N_{\{n\} \{m\} } &=& \int d\mu (U) ~ \Psi^*_{\{n\}}(U)\,  \Psi_{\{m\}}(U)  =
 \int d\mu (U) ~ \bra{\{n\}} U \ket{0}\, \bra{0} U^\dagger \ket{\{m\}} \nonumber\\
 &\equiv& \bra{ \{ n\} } N \ket{ \{m \} } ,
 \label{18}
 \eeqar
 where $N$ denotes the operator
 \beq
 N =  \int d\mu (U) ~ U \ket{0} \bra{0} U^\dagger .
\label{18a}
\eeq
Complex conjugation of this equation shows that $N_{\{n\} \{m\} } $ is a hermitian matrix
or it can be viewe as a hermitian operator on the states  (\ref{13}) of the Verma module.
Further by translational invariance of the integration measure we have
\beqar
N_{\{n\} \{m\} }&=&  \int d\mu (V_L U) ~ \bra{\{n\}} V_L U \ket{0}\, \bra{0} U^\dagger V_L^\dagger \ket{\{m\}} \nonumber\\
&=& \int d\mu (U) ~ \bra{\{n\}} V_L U \ket{0}\, \bra{0} U^\dagger V_L^\dagger \ket{\{m\}} .
\label{19}
\eeqar
We take $V_L$ to be of the form
\beq
V_L = \exp\left[ \sum_k  {\bar \theta}_{k } L_k  - \theta_k \, L_{-k} \right] ,
\label{20}
\eeq
where $\theta_k$, ${\bar \theta}_k$ are infinitesimal parameters
which are constant, i.e., independent of $s_l$, $\bs_l$.
To linear order in $\theta_k$, equation (\ref{19}) then leads to
\beq
\int d\mu(U)\, \left[ \bra{\{n\}} L_{-k} U \ket{0} \bra{0} U^\dagger \ket{\{m\}}
- \bra{\{n\}}  U \ket{0} \bra{0} U^\dagger L_{-k} \ket{\{m\}} \right] = 0 .
\label{21}
\eeq
Since we can write $\bra{\{n\}} L_{-k} = (L_{-k})_{\{n\} \{m\} } \bra{\{m\}}$, this translates to
$[ L_{-k} , N] = 0$. Similarly, from the coefficient of
${\bar\theta}_\bn$, we also get $[L_k , N] = 0$. It is also easy to see that
$[L_0, N] = 0$. 
Since $N$ commutes with all $L_k$ for all states of the form (\ref{13}),
and since these states form a basis, we can write
$N_{\{n\} \{m\} } = \Lambda \, \delta_{\{n\} \{m\} } $, where $\Lambda$ is a constant
independent of the state labels $\{n\}$, $\{m\}$.
But $\Lambda$ can depend on $c$ and $h$ which characterize the representation.
This constant $\Lambda$ can be absorbed into
the definition of the measure $d\mu$; we will do this from now on, so that
we have the result
\beq
\int d\mu (U) ~ \Psi^*_{\{n\}} (U) \,  \Psi_{\{m\}}(U)   = \delta_{\{n\} \{m\} }  .
\label{22}
\eeq
Given the definition of the wave functions (\ref{14}), this can also be viewed as
the completeness relation
\beq
\int d\mu(U)~ U \ket{0} \bra{0} U^\dagger = {\mathbb 1} .
\label{23}
\eeq

We now come to the question of whether $\Lambda$ is finite , i.e., whether
the normalization integrals exist. From the discussion given above,
$\Lambda$ is given by the integral of $e^{-K}$,
\beq
\Lambda = \int d\mu \bra{0} U\ket{0} \, \bra{0} U^\dagger \ket{0}
= \int d\mu \, e^{-K} 
\label{23a}
\eeq

There are two issues one needs to address here. There are an infinite
number of coordinates, so a suitable regularization has to be used, 
as is the case for the functional integral in any field theory.
We have to assume the existence of such a regularization to make the question well-defined.
 The simplest possibility would be to truncate the integral to a finite number of
variables, say $N$, taking the limit $N \rightarrow \infty$ after 
the expectation values of operators have the been evaluated.
Even after truncation to a finite set of modes,
there is still the question of whether the integration
over each coordinate is convergent. In other words, does the factor
$e^{-K}$ provide sufficient damping for large magnitudes 
for the coordinates? 
We will now present three sets of arguments
pointing out features of $K$ relevant to these questions.

First of all, we note that in
the $SL(2, {\mathbb R})$ case, the integral of $e^{-K}$ does exist, 
 with the restriction $h > {\half}$, as indicated in section 2.
 In particular, $K \rightarrow \infty, \, e^{-K} \rightarrow 0$ as $\vert w\vert 
 \rightarrow \infty$ (which is equivalent to the limit 
 $\vert s\vert \rightarrow 1$).
 Therefore, as the first step towards analyzing the asymptotic behavior of $K$, we can use $SL(2, \mathbb{R})$ subalgebras,
defined by $L_0, L_{\pm m}$, for fixed $m$.
If we consider all $(w_n, \bw_n)$ to be zero except for one pair, say,
$(w_m, \bw_m)$, then $K$ will reduce to the K\"ahler potential
for $SL(2, \mathbb{R})$ and hence we get
$K \rightarrow \infty$, $e^{-K} \rightarrow 0$ as
$\vert w_m\vert \rightarrow \infty$. Thus for every plane
in the orbit space, $K \rightarrow \infty$ for large $\vert w\vert$'s.
This property also holds if we take $\vert w_m \vert \rightarrow
\infty$ holding all other $w$'s fixed and finite, but not necessarily zero.
Towards this, 
 notice that $U$ involves the combination
  $\bw_n L_n - w_n L_{-n}$. We then write
  \beq
  \sum \bw_n L_n = \sqrt{2}\, \bw_2 \L_2,
  \hskip .2in
  \L_2 = {1\over \sqrt{2}} \left( L_2 + \sum_{n\neq 2} {\bw_n L_n \over \bw_2}\right).
  \label{23b}
  \eeq
This is for the case where we plan to take $\vert w_2\vert \rightarrow \infty$ keeping other $w$'s fixed.
  The Virasoro algebra leads to the commutation rules
  \beqar
  [\L_2, \L_{-2} ]&=& 2\, \L_0 + \mathbb{X}\nonumber\\
  {}[\L_0 , \L_{2} ] &=& - 2\, \L_{ 2} - \mathbb{Y}
  \label{23c}\\
  {}[\L_0 , \L_{-2} ] &=&  2\, \L_{ -2} + {\mathbb{Y}}^\dagger
    \nonumber
  \eeqar
  where $\L_0 = L_0 + (c/8)$ and
  \beqar
 \mathbb{X}&=& {1\over 2} \sum_{n\neq 2}\left[
  {(n+2) w_n \over w_2} L_{2-n} +   {(n+2) \bw_n \over \bw_2} L_{n-2} 
  +  {c \over 12} {\bw_n w_n \over \bw_2 w_2} (n^3-n)\right]\nonumber\\
  && +
 {1\over 2}\sum_{n,m\neq 2}  {(n+m)  \bw_n w_m \over \bw_2 w_2} L_{n-m}
  \nonumber\\
  \mathbb{Y}&=&  \sum_{n\neq 2} {(n+2)\bw_n\over \sqrt{2} \bw_2}\, L_n 
  \label{23d}
  \eeqar
  The terms $\mathbb{X}$, $\mathbb{Y}$ vanish as $\vert w_2\vert
  \rightarrow \infty$ keeping other coordinates finite,
leading to an $SL(2, \mathbb{R})$ subalgebra
of $\L_0$, $\L_{\pm 2}$. Notice also that we can write
\beq
U = \exp\left( \sum_n \bw_n L_n - w_n L_{-n} \right)
= \exp\left( \sqrt{2}(\bw_2 \L_2 - w_2 \L_{-2} )\right)
\label{23e}
\eeq
In calculating $\bra{0} U \ket{0}$, $\bra{0} U^\dagger \ket{0}$, we use the
Baker-Campbell-Hausdorff formula to rearrange
$U^\dagger$ as
\beq
U^\dagger =  e^{ f_n L_{-n} } e^{- {\half} (W^0 L_0 + W )}
e^{{\bar f}_n L_n}
\label{23f}
\eeq
This rearrangement only uses the commutation rules.
Since the structure constants in the commutation rules (\ref{23c}) become those
of $SL(2, \mathbb{R})$ as $\vert w_2\vert \rightarrow \infty$,
keeping all other $w_n$'s fixed, we get the result
\beq
K = K_{SL(2, \mathbb{R})} (\sqrt{2} w_2, \sqrt{2}\bw_2)
+ \cdots
\label{23g}
\eeq
where the ellipsis indicates terms which vanish in the limit.
It then follows that $e^{- K}$ vanishes in the limit
$\vert w_2\vert \rightarrow \infty$, keeping all other $w_n$'s fixed.
A similar argument holds for any $\vert w_m\vert \rightarrow \infty$
keeping all other $w$'s fixed.

Our second observation relates to the behavior of $K$ under scaling, i.e., how it behaves as we go to large $\vert w_n\vert$ uniformly for all
$n$. We can see that a common scaling up of the $w$'s 
will increase $K$. 
Writing $U= e^{i C}$ in terms of the hermitian
operator $C = - i \sum_n (\bw_n L_n - w_n L_{-n} )$,
we find
\beqar
\exp \left[ -  K ((1+\epsilon ) w, (1+ \epsilon )\bw )\right]
&=& \bra{0} e^{- i (1+\epsilon ) C }\ket{0} \,\bra{0} e^{ i (1+\epsilon ) C }\ket{0}\nonumber\\
&=& \exp \left[ - K (w, \bw )\right]
- i \epsilon \bra{0} e^{- i C } C \ket{0} \bra{0} e^{i C } \ket{0}\nonumber\\
&&~+ \, i \epsilon \bra{0} e^{ i C } C \ket{0} \bra{0} e^{-i C } \ket{0} + \cdots
\label{23h}
\eeqar
This leads to
\beqar
\sum_n\left[ w_n {\del \over \del w_n } + \bw_n {\del \over \del \bw_n} 
\right] K &=&  2\, e^{ K} \, \Bigl[ \bra{0} \cos C \ket{0}
\bra{0} C \sin C \ket{0}\nonumber\\
&&\hskip .2in - \bra{0} \sin C \ket{0} \bra{0} C \cos C \ket{0}
\Bigr]
\label{23i}
 \eeqar
 where, for brevity, we use
 $\la \cos C\ra = \bra{0} \cos C \ket{0}$, $\la C \sin C\ra
 = \bra{0}  C\sin C \ket{0}$, etc.
Since $C$ is hermitian, we can diagonalize it. If $\ket{\alpha}$
denote the states which diagonalize $C$, with eigenvalues
$c_\alpha$, we can write
\beqar
\la\cos C \ra
\la C \sin C \ra - \la \sin C \ra \la C \cos C \ra
&=& \sum p_\alpha p_\beta \, c_\beta \left[ \cos c_\alpha \sin c_\beta
- \sin c_\alpha \cos c_\beta \right]\nonumber\\
&=&-  \sum p_\alpha p_\beta \, c_\beta \sin (c_\alpha - c_\beta)\nonumber\\
&=&{1\over 2}  \sum p_\alpha p_\beta (c_\alpha - \, c_\beta) \sin (c_\alpha - c_\beta)
\label{23j}
\eeqar
where $p_\alpha = \vert \braket{0|\alpha}\vert^2$.
This shows that the right hand side of (\ref{23i}) is always positive.
Therefore $K$ will continually increase with $\lambda$ under scaling 
 $(w_n, \bw_n) \rightarrow (\lambda w_n, \lambda \bw_n) $,
$\lambda > 0$.
 The limits of integration for the $w$'s should be defined by
 the vanishing of $e^{- K}$ in (\ref{23a}), and we see that, with the
 two properties of $K$ given above, 
 we can expect convergence for integration over each mode.
 
 Our third observation is about regularizing the integration
 over the infinite number of modes.
 As mentioned before, a regularization truncating to a finite number
 of coordinates, say $N$, is needed, with
 the limit $N \rightarrow \infty$ eventually.
 Ultimately, this will require treating $K$ as if it is the action for
 a quantum field theory. Regularization by truncation to
 a finite number of modes is familiar from field theory.
This is also somewhat similar to
 what is done in \cite{S-W}, 
which presents the calculation of the integral $\int d \mu ~ e^{- H}$,
where $d\mu$ is the symplectic measure on {{$\widehat{{\rm diff}\,S^1}$}} /$SL(2, {\mathbb R})$
and $H$ is the Hamiltonian for translations on $S^1$, i.e., for the
action of $L_0$.
The result up to two loops is obtained; regularization is implicit
as in standard field theory calculations.
It is also argued that
 a similar result holds for {{$\widehat{{\rm diff}\,S^1}$}}/$U(1)$, which is the orbit we are interested in.
For the present situation, it is not the integral of $e^{-H}$ we need, but
notice that we do have a field theoretic way of
understanding the normalization integral (\ref{23a}).
The action is not defined by the Hamiltonian corresponding to
the action of $L_0$, but rather it is given by
the K\"ahler potential $K$. From 
 (\ref{KP6}) we see that $K$ can indeed be viewed as an action
 for a complex field $\chi$.
 In particular, if we consider the large $c$ limit, then we can
 scale $\chi \rightarrow \chi/\sqrt{c}$ for all the modes
 $s_n$, $\bs_n$, $n\geq 2$. The set of terms in $K$ which are
 proportional to $c$ do not
 involve $s_1$, $\bs_1$, so this scaling does not affect
 those modes. For large $c$, we then get
 \beqar
 K &\approx& {1\over 4\pi} \int d\tau\, \left[ {i\over 12} ({\bar \chi}'' \chi' - \chi''
 {\bar \chi}' ) + {i \over 6} {\bar \chi} \chi' - 4 h \log (1- \bs_1 s_1)
 \right]\nonumber\\
 &=& {1\over 12}\sum_n (n^3 -n) \bs_n s_n - 2 h \log (1- \bs_1 s_1)
 \label{23k}
 \eeqar
We get a Gaussian integral for $s_n$, $n \neq 1$ and
the $SL(2, \mathbb{R})$ integral for the $s_1$ mode.
The determinant arising form the Gaussian term can be regularized
as is done in any field theory, for example, using the
a cutoff or using the $\zeta$-function.
We see that the integral does exist in the sense of quantum field theory,
at least for large $c$ (and for $h >{\half}$).

Another observation of relevance would be that, just based on algebraic considerations, the states of the Verma module do
have a finite norm, and this, in our language, is related to
to the integral in (\ref{18}).
Further, Virasoro representations occur in the partition function for
(2+1)-dimensional gravity \cite{{witten2},{maloney}}.
Defining such partition functions is, in an indirect way,
equivalent to the existence of $\Lambda$.

We also note that, bypassing the normalization integral in
(\ref{18}), it may be possible to 
obtain a resolution of identity by suitable restrictions, either by going to a
quotient space \cite{calix}, or by considering a family of coadjoint orbits \cite{isham}, or by considering a subset of
coherent states \cite{bievre}.\footnote{I thank a reviewer of an early version of this paper for pointing this out and bringing the 
relevant papers to my attention.}

\section{Symbols and star products}

We are now in a position to
define the symbols associated to an operator and the corresponding star products.\footnote{In the terminology often used in the context of Berezin-Toeplitz quantization \cite{BT}, these are covariant symbols .}
We consider operators $A$ $B$, etc. acting on the states (\ref{13}). We may also view them as matrices
with elements of the form
$A_{\{n\} \{m\} } $, $B_{\{n\} \{m\} } $, etc.
The symbols corresponding to these
operators will be defined as
\beqar
(A) &=& \sum_{\{n\},\{m\}}\bra{0} U^\dagger \ket{\{n\}} A_{\{n\} \{m\} } \bra{\{m\}} U \ket{0}
= \bra{0} U^\dagger A U \ket{0} \nonumber\\
(B) &=&  \bra{0} U^\dagger B U \ket{0} .
\label{24}
\eeqar
The symbol corresponding to a product of two operators takes the form
\beq
(AB) = \bra{0} U^\dagger A B U \ket{0} .
\label{25}
\eeq
This can be written in terms of the symbols of the individual operators
and the their derivatives which will constitute the star product.
For this, we first note that the completeness relation for the states
(\ref{11}) takes the form
\beq
\sum_{\{n \}, \{ m \} }  \ket{\{ {\tilde n}\} } (\M^{-1})_{\{ n\}, \{m \} } \, \bra{\{ {\tilde m}\} }
= {\mathbb 1} .
\label{26}
\eeq
We can now rewrite the symbol of the operator product in (\ref{25}) by using the completeness
relation as
\beqar
(AB) &=& \bra{0} U^\dagger A \,U {\mathbb 1} U^\dagger\,B U \ket{0}\nonumber\\
&=&\sum_{\{n\}, \{ m \} } \bra{0} U^\dagger A U  
 \ket{\{ {\tilde n}\} } (\M^{-1})_{\{ n\}, \{m \} } \, \bra{\{ {\tilde m}\} } U^\dagger\,B U \ket{0}
 \nonumber\\
 &=&\sum_{\{n\}, \{ m \} } \bra{0} U^\dagger A U  
  \cdots  L_{-2}^{n_2} L_{-1}^{n_1} \, \ket{0}
  (\M^{-1})_{\{ n\}, \{m \} }  
  \bra{0} L_1^{m_1} L_2^{m_2}  \cdots U^\dagger\,B U \ket{0}\nonumber\\
  &=& \bra{0} U^\dagger A U  \ket{0}\bra{0} U^\dagger\,B U \ket{0}
  + \bra{0} U^\dagger A U L_{-1} \ket{0} {1\over 2 h} \bra{0} L_1 U^\dagger\,B U \ket{0}
+ \cdots\nonumber\\
&\equiv& (A) * (B) .
\label{27}
\eeqar
Terms of the form $\bra{0} U^\dagger A U L_{-k} L_{-n}\ket{0}$ can be simplified using the following relations which are a consequence of
equations (\ref{4}) and (\ref{9}):
\beqar
{\del U \over \del s_k} &=& -\sum_n \E^n_k \, U L_{-n} + U \left( \E^0_k L_0 + \E_k {\mathbb 1} \right)
\nonumber\\
{\del U^\dagger \over \del s_k} &=&   \sum_n\E^n_k \,  L_{-n} U^\dagger -  \left( \E^0_k L_0 + \E_k {\mathbb 1} \right) 
U^\dagger
\label{28}\\
{\del U \over \del \bs_k} &=&  \sum_n \bcalE^n_k U  L_n - U \left( \bcalE^0_k L_0 + \bcalE_k {\mathbb 1}
\right)\nonumber\\
{\del U^\dagger \over \del \bs_k} &=& - \sum_n \bcalE^n_k L_n \, U^\dagger
+ \left(\bcalE^0_k L_0 + \bcalE_k {\mathbb 1} \right) U^\dagger .
\label{29}
\eeqar
We now define covariant derivatives $\D_n$, ${\bar\D}_n$ by
\beqar
\D_n&=& - \sum_k ( \E^{-1})^k_n \left[ {\del \over \del s_k} - {\ell}_0\, \E^0_k \right]\nonumber\\
{\bar\D}_n &=& - \sum_k (\bcalE^{-1})^k_n \left[ {\del \over \del \bs_k} +
{\ell}_0\, \bcalE^0_k \right] ,
\label{30}
\eeqar
where $\ell_0$ denotes the eigenvalue of $L_0$ for the expression
on which these covariant derivatives act; i.e., it is the level number
of the expression to the right of these derivatives.
Using (\ref{28}, \ref{29}), it is easy to check that we can write
\beqar
\D_{\{n\}} (A) \equiv\cdots \D_{n_2} \D_{n_1} (A) &=&\bra{0} U^\dagger A U \cdots L_{-n_2} L_{-n_1} \ket{0}\nonumber\\
{\bar\D}_{\{n\}} (B) \equiv \cdots {\bar \D}_{n_2} {\bar \D}_{n_1} (B) &=&
\bra{0} L_{n_1} L_{n_2} \cdots U^\dagger B U \ket{0} .
\label{31}
\eeqar
The star product can thus be rewritten as
\beq
(A)*(B) = \sum_{\{n\}, \{ m \} } \D_{\{n\}} (A) \,
  (\M^{-1})_{\{ n\}, \{m \} }  \,
{\bar \D}_{\{ m\}} (B) .
\label{32}
\eeq
The symbols themselves have a value of zero for $\ell_0$; we also have
$\ell_0 = \pm n$ for 
$\D_n$ and ${\bar\D}_n$, respectively.

The second term on the right hand side in the expansion in (\ref{27}), which involves only 
$L_\pm1$, comes
from the $SL(2, {\mathbb R})$ subalgebra. In fact, as argued below, if we consider just the states generated by powers of $L_{-1}$, the star product in
(\ref{32}) will become the star product for $SL(2, \mathbb{R})$ states
defined in section 2.

It is useful to work out the next term which is at level 2.
The matrix of inner products and its inverse are given by
\beqar
\M &=& \left[ \begin{matrix} 
8 h^2 + 4 h&6 h\\ 6h& 4 h + {\half c} \\ \end{matrix} \right],
\hskip .2in
\M^{-1} = {1\over \det \M} \left[ \begin{matrix} 
4 h + {\half} c& - 6 h\\ - 6h& 8 h^2 + 4 h \\ \end{matrix} \right]\nonumber\\
\det \M&=& 2  h  (2 h +1) (c -1) + 2 h (4 h-1)^2 .
\label{33}
\eeqar
Here the matrix elements refer to the states
$\ket{1} = L_{-1}^2 \ket{0}$, $\ket{2} = L_{-2} \ket{0}$.
Explicitly, to this order, we get
\beqar
(A)* (B) &=& (A) (B) + {1\over 2 h} (\E^{-1})^k_1 {\del (A) \over \del s_k}
\,(\bcalE^{-1})^{k'}_1 {\del (B) \over \del \bs_{k'}}\nonumber\\
&&
+ { 4 h +  {\half} c \over \det \M}
\left[ \left((\E^{-1})^k_1 {\del \over \del s_k}  - (\E^{-1})^k_1\E^0_k \right)  (\E^{-1})^l_1 {\del (A) \over \del s_l}\times\right.\nonumber\\
&&\hskip .8in \left.
\left( (\bcalE^{-1})^{k'}_1 {\del \over \del \bs_{k'}}  - 
(\bcalE^{-1})^{k'}_1\bcalE^0_{k'} \right)  (\bcalE^{-1})^{l'}_1 {\del (B) \over \del \bs_{l'}} \right]\nonumber\\
&&+ {6 h \over \det\M} \left[  \left((\E^{-1})^k_1 {\del \over \del s_k}  - (\E^{-1})^k_1\E^0_k \right)  (\E^{-1})^l_1 {\del (A) \over \del s_l}
\, (\bcalE^{-1})^{l'}_2 {\del (B)\over \del \bs_{l'}}\right.\nonumber\\
&&\hskip .8in \left.+ (\E^{-1})^k_2 {\del (A)\over \del s_k}
\left( (\bcalE^{-1})^{k'}_1 {\del \over \del \bs_{k'}}  - 
(\bcalE^{-1})^{k'}_1\bcalE^0_{k'} \right)  (\E^{-1})^{l'}_1 {\del (B) \over \del \bs_{l'}} \right]\nonumber\\
&&\hskip .8in + {8 h^2 + 4 h \over \det \M}
\left[ (\E^{-1})^k_2 {\del (A)\over \del s_k}
(\bcalE^{-1})^{l'}_2 {\del (B)\over \del \bs_{l'}}\right] + \cdots
\label{33a}
\eeqar
There is summation over $k$, $k'$, $l$, $l'$ in various expressions in this equation.

If we take the large $c$ limit at fixed $h$, the matrix $\M^{-1}$ reduces to its
$(1,1)$-component
$\M^{-1}_{11} \approx (8 h^2 + 4 h)^{-1}$ with all other elements zero.
The corresponding term, along with the terms with first derivatives of $(A)$ and $(B)$ given in the first line, give the first two  terms of
the star-product for 
the orbit $SL(2, {\mathbb R})/U(1)$ labeled by the highest weight $L_0 = h$.

It is easy to see that this property is obtained for higher terms as well. 
For this, consider the matrix $\M$ at level $n$.
The element $\M_{11}$ arises from 
$L_{-1}^n \ket{0}$, so this term is exactly what it is for 
$SL(2, \mathbb{R})$. Further, terms arising from
$L_{-2} L_{-1}^{n-2}\ket{0}$, $L_{-3} L_{-1}^{n-3}\ket{0}$,
$L_{-3} L_{-2} L_{-1}^{n-5}\ket{0}$, etc. will all have powers
of $c$, since we get the central terms for
$\bra{0} L_k L_{-k} \ket{0}$. The principal minor or cofactor corresponding to
$\M_{11}$ thus dominates the determinant as $c\rightarrow \infty$, and is order $c^{n-1}$,
while the cofactors corresponding to the other elements will have
smaller powers of $c$. As a result, in the inverse of $\M$, the element
$(\M^{-1})_{11}$ dominates as $c \rightarrow \infty$,
with the limiting value $1/\M_{11}$, while
other elements tend to zero. 
Since $1/\M_{11}$ corresponds to $SL(2, \mathbb{R})$, we get the result
that the star product for the Virasoro group characterized by
$h$, $c$ becomes the star product for the $SL(2, {\mathbb R})/U(1)$ orbit labeled
by $h$; i.e.,
\beq
(A)* (B) \Bigl]_{{\rm Virasoro}, h, c} \hskip .1in \longrightarrow  \hskip .1in
(A)*(B)\Bigl]_{SL(2, {\mathbb R}), h}\hskip .2in{\rm as} ~ c \rightarrow \infty .
\label{34}
\eeq
This reduction also conforms to what is expected from
the large $c$ behavior discussed in connection with (\ref{23k}).
In the context of this result, it may be interesting to recall
 that the $c \rightarrow \infty$ limit is important in the context of
semiclassical limits of partition functions for (2+1)-dimensional gravity \cite{{witten2},{maloney}}.
The coset $SL(2, {\mathbb R})/U(1)$ with the star product 
in (\ref{34}), is the noncommutative version of ${\rm AdS}_2$. For more on this
space, see \cite{ads2}.

\section{Discussion}

We have worked out the construction of coherent states for the Virasoro group using a class of orbits with $c > 1$, $h > {\half}$. 
The basic results are in section 3.  
The wave functions for the coherent states are given in
(\ref{14}). They satisfy a certain antiholomorphicity property, as is evident
from (\ref{16d}). We also give an explicit formula for the K\"ahler potential
$K$ in terms of complex fields defined on the circle.
We consider some of the key properties of $K$ and also discuss
the normalization integral for the wave functions as
the partition function for the one-dimensional
field theory corresponding to this $K$.

In the case of the well-known coherent states on orbits of compact
groups as well as the historically first case of the oscillator, one can define
reproducing kernels via the overlap of coherent state wave functions.
It would be interesting to explore the properties of such kernels in
the Virasoro case. 

Also, we have restricted ourselves to $c> 1$,
since we were partly motivated by possible applications
to (2+1)-dimensional gravity. However, 
the discrete set of unitary $c< 1$ representations are also very interesting from a physics
point of view, since they are relevant to the so-called minimal models in CFT.
Coherent states for such cases are also worth exploring,
but are obviously beyond the scope of this work.

Turning to the more physical side
of things, our results can be viewed as furnishing a 
noncommutative version of the infinite dimensional K\"ahler space
{{$\widehat{{\rm diff}\,S^1}$/$S^1$}}. The star product (\ref{32}) gives the 
noncommutative algebra for functions.
As mentioned in the Introduction,
this analysis was partially motivated by potential application to 
(2+1)-dimensional gravity.
In the spirit of noncommutative geometry, one can use a Hilbert space
of states to model the spatial manifold \cite{{NC}, {CC}}.
We have recently argued for elaborating this framework, with a doubling of the Hilbert
space as in thermofield dynamics, with the gauge fields for gravity, i.e., the frame fields and the
spin connection, coupling to the two Hilbert spaces in parity-conjugate ways
\cite{Nair}.
In this framework, it is possible to obtain the Einstein-Hilbert action
for gravity in (2+1) dimensions (with a nonzero cosmological constant), upon taking
the commutative limit. The explicit calculations were done
using quantization of the orbits $SU(2)/U(1)$ or $SL(2, {\mathbb R})/ U(1)$
to model the corresponding noncommutative spaces.
However, since the partition function for gravity naturally involves representations of the 
Virasoro algebra due to the contributions from black holes \cite{{witten2},{maloney}},
one can ask whether it is possible to extend the analysis and
use the carrier space of the representation
as the Hilbert space of interest modeling the noncommutative space.
The coherent states discussed here provide a way to define symbols and star products 
for such a formulation.
We should expect that the commutative limit, which is the large $(c, h)$ limit, 
will then lead to
the Einstein-Hilbert action.

Our results may also be interpreted in terms
of a mock quantum Hall system. We have analyzed
the quantum Hall problem in arbitrary dimensions in a series of papers
and argued that the lowest Landau level of such systems
model the corresponding noncommutative spaces \cite{KN}.
Effective actions, edge states, etc. were analyzed in such cases.
The present discussion may be viewed as another example of this,
now applied to an infinite-dimensional case.

As noted before, the K\"ahler potential defines an interesting
one dimensional field theory in its own right, given by
\begin{align}
Z =& \int d\mu \, e^{-K}\label{Dis1}\\
K =& {1\over 4\pi} \int_0^{2\pi}  \!\!\!d\tau\,
 \Biggl\{ {c \over 12} \left[
 {i \over (1+ \chi' +{\bar\chi}')}\left( { {\bar\chi}'' \chi' \over (1+ {\bar\chi}')}
- {\chi'' {\bar \chi}' \over (1+ \chi')} \right)\right]\nonumber\\
&\hskip .7in + \left( 4 h - {c\over 6}\right) (- i \,{\bar \chi}\, \chi' )
 - 4 h \log(1- i {\bar\vf}' \vf) \Biggr\}
\nonumber
\end{align}
This should prove to be an interesting theory since it is closely tied to the
Virasoro group. We propose to investigate this further in future.

\vskip .15in
I thank Edward Witten for a useful comment in relation to reference
\cite{S-W}. I also express my gratitude
to Charles Nash and Denjoe O'Connor for
many useful discussions. Some of this work was done during a visit to
the Dublin Institute for Advanced Studies (DIAS).
I also thank Denjoe O'Connor and members
of the School of Theoretical Physics at DIAS for their warm hospitality.

This research was supported in part by the U.S.\ National Science
Foundation grant PHY-2112729.

\section*{Appendix}
\def\theequation{A\arabic{equation}}
\setcounter{equation}{0}

In this Appendix we will go over some of the details of the results mentioned in section 3.
\vskip .05in\noindent
\underline{Analysis near the origin}

The quantities $w_n$, $\bw_n$ in $U$ will be chosen as functions of the coordinates
$s_n$, $\bs_n$, i.e.,
\beq
\bw_k = \bs_k + \bw_k^{(2)} + \bw_k^{(3)} + \cdots ,
\label{A0}
\eeq
where $\bw_k^{(r)}$ is of order $r$ in powers of $s_n$, $\bs_n$.
Our strategy is to write down $U^{-1} dU$ and choose
these functions such that $\E^n$ is a holomorphic differential
and $\bcalE^n$ is an antiholomorphic differential.
From the definition of $U$ in (\ref{3}), we can write
\beqar
U^{-1} d U &=& \sum_n \int_0^1 d\alpha~
e^{-\alpha X} \left( d \bw_n L_n - d w_n L_{-n} \right) e^{\alpha X}\nonumber\\
X&=& \sum_m (\bw_m L_m - w_m L_{-m} ) .
\label{A1}
\eeqar
Expanding the exponential, we can write out the first two terms
as
\beqar
{\rm Term}~ 1&=& \sum d\bw_n  L_n - dw_n L_{-n}\nonumber\\
{\rm Term}~2&=& - {1\over 2!}\sum_n  [ X, d\bw_n L_n - dw_n L_{-n}]\nonumber\\
&=& - {1\over 2!} \sum_{m,n}\Bigl[ (m-n) \bw_m d\bw_n L_{m+n}
-(m-n) w_m dw_n L_{-m-n} \label{A2}\\
&&\hskip .1in + (m+n) ( w_n d\bw_m - \bw_m dw_n ) L_{m-n}
+ {c \over 12} (n^3 -n) (w_n d\bw_n - \bw_n dw_n ) \Bigr] .\nonumber
\eeqar
The $L_{m-n}$ can have terms of the form $L_k$ and $L_{-k}$, $k > 0$, as well as
$L_0$ terms. Separating these out
we find
\beqar
{\rm Term}~2&=& - {1\over 2!} 
\sum_n (w_n d\bw_n - \bw_n dw_n ) \left( 2 n L_0 + {c\over 12} (n^3 - n){\mathbb 1} \right)\nonumber\\
&&-{1\over 2!}\sum_k \left[ {\bar C}^{(2)}_k L_k  - C^{(2)}_k L_{-k} \right] ,\nonumber\\
{\bar C}^{(2)}_k &=& \sum_{n=1}^{k-1} (k -2 n) \bw_{k-n} d \bw_n\, \Theta(k-3)\nonumber\\
&&
- \sum_n (k +2 n) ( \bw_{k+n} d w_n - w_n d \bw_{k+n} )\, \Theta(k-1) ,
\label{A3}
\eeqar
with ${C_k^{(2)}} $ being the complex conjugate of ${\bar C}^{(2)}_k$. Also
$\Theta(k- a) = 1$ for $k \geq a$ and zero otherwise. From the expression for
${\bar C}^{(2)}_k$, we see that there is one term
in the coefficient of $L_k$ which has $dw_n$. We rewrite this term using
\beq
\sum (k+ 2n) \bw_{k+n} dw_n = d \left[ \sum (k + 2n ) \bw_{k+n} w_n \right] -
\sum (k + 2n) w_n d\bw_{k+n}  .
\label{A4}
\eeq
Thus we find
\beqar
{\rm Coefficient~of~} L_k &=&
d \left[ \bw_k + {1\over 2} \sum (k+ 2n)  \bw_{k+n} w_n \right]
- {1\over 2} \sum_1^{k-1} (k - 2n ) \bw_{k-n} d\bw_n \, \Theta (k-3)\nonumber\\
&&- \sum_n ( k + 2n) w_n d\bw_{k+n} \, \Theta( k-1) .
\label{A5}
\eeqar
We now define $w_k$, $\bw_{k}$ as functions of complex coordinates
$s_k$, $\bs_k$, with an expansion around the origin as
\beq
\bw_k = \bs_k - {1\over 2} \sum_n ( k+ 2 n) \,\bs_{k + n} s_n
+ {\bar w}^{(3)}_k + \cdots ,
\label{A6}
\eeq
where ${\bar w}^{(3)}_k$ denote terms which are cubic
in $s_n, \bs_n$.
The coefficient of $L_k$ now becomes
\beqar
{\rm Coefficient~of~} L_k &=& d\bs_k - {1\over 2} \sum_1^{k-1} ( k - 2 n) 
\bs_{k - n} d\bs_n\, \Theta(k-3) \nonumber\\
&&- \sum (k + 2n) s_k d\bs_{k + n}\, \Theta(k-1)
+ \cdots\nonumber\\
&\equiv& d\bs_k - {1\over 2} \sum_l D^{(2)}_{k l} d\bs_l ~+ \cdots .
\label{A7}
\eeqar
We see that, to this order, there are no $ds_n$-terms in the coefficient of $L_k$, $k>0$; i.e.,
it is an antiholomorphic one-form. To the same approximation, the coefficient
of the $L_0$ and central terms become
\beq
(L_0, {\mathbb 1} )\mhyphen{\rm terms} =
 - {1\over 2!} 
\sum_n (s_n d\bs_n - \bs_n ds_n ) \left( 2 n L_0 + {c\over 12} (n^3 - n){\mathbb 1} \right) .
\label{A8}
\eeq

The next term in the expansion of (\ref{A1}), corresponding to the double commutator of
$X$ with $\sum d\bw_n  L_n - dw_n L_{-n}$ is of the form
\beq
{\rm Term~} 3 = {1\over 3!} \left[ 
\sum ( w_n {\bar C}^{(2)}_n - \bw_n C^{(2)}_n ) \bigl[ 2 n L_0 + {c\over 12} (n^3 -n )\bigr]
+ \sum {\bar C}^{(3)}_k L_k - C^{(3)}_k L_{-k} \right] .
\label{A9}
\eeq
We can use $w_n \approx s_n$, $\bw_n \approx \bs_n$ in working out the 
cubic terms in this expression, to get expressions valid to the third order
in $s_n$, $\bs_n$.
To this order, ${\bar C}^{(3)}_k$ is given by
\beqar
{\bar C}^{(3)}_k &\approx& \sum_n 2 k n ( s_n d\bs_n - \bs_n ds_n ) \bs_k
+ \sum \delta_{k, m+n} (m -n) \bs_m {\bar C}^{(2)}_n\nonumber\\
&&+ \sum \delta_{k, m-n} (m+n) \left[ s_n {\bar C}^{(2)}_m - \bs_m C^{(2)}_n
\right] .
\label{A9a}
\eeqar
In this expression, a term like $\bs_k \bs_n ds_n$ (which is a holomorphic
form and hence not what we want) can be removed by
a term like $\bs_k \bs_n s_n$ in the expression for ${\bar w}^{(3)}_k$.
The simplification of the expression for ${\bar C}^{(3)}_k$
is straightforward but long. With some rearrangements of terms, it
can be brought to the form
\beq
{\bar C}^{(3)}_k = \sum_l D^{(3)}_{k,l} d\bs_l - d \chi^{(3)}_k 
- {3\over 2}\sum \delta_{k, r-s-n} (r+s+n) (n-s) \bs_m s_n ds_s .
\label{A9b}
\eeq
The expression for $D^{(3)}_{k,l} d\bs_l $ is long and not important for our argument
(since it is an antiholomorphic form anyway),
but we give it here for the sake of completeness,
\beqar
\sum_l D^{(3)}_{k,l} d\bs_l &=&\sum\Bigl[  s_n \bs_k d\bs_n + 2 k n \bs_n s_n ds_k
+ \delta_{k, m+n} \delta_{n, r+s} (m-n) (r-s) \bs_m \bs_r d\bs_{s}\nonumber\\
&&+ 2\, \delta_{k, m+n} \delta_{n, r-s} (m-n) (r+s) \bs_m s_s ds_r\nonumber\\
&&+ \delta_{k, m+n} \delta_{n, r-s} (m-n) (r+s) \bs_r s_s d\bs_m\nonumber\\
&&+ \delta_{k, m-n} \delta_{m, r+s} (m+n) (r-s) s_n \bs_r d\bs_{s}\nonumber\\
&&+ \delta_{k, m-n} \delta_{m, r-s} (m+n) (r+s) s_n s_s d\bs_r\nonumber\\
&&+ {\half} \delta_{k, r-s-n} \sigma(r,s,n) s_n s_s d\bs_r
+ 2 \delta_{k, m-n} \delta_{n, r-s} (m+n) (r+s) \bs_m s_r d\bs_{s} \nonumber\\
&&+ \delta_{k, m-n} \delta_{n, r-s} (m+n) (r+s) \bs_{s} s_r d\bs_m
\Bigr] .
\label{A9c}
\eeqar
Also $\chi_k$ is given by
\beqar
\chi^{(3)}_k &=& \sum\Bigl[ 2 k n \bs_n \bs_k s_n + \delta_{k, m+n} \delta_{n, r-s}
(m-n) (r+s) \bs_m \bs_r s_s \nonumber\\
&&+ {\half} \sigma (r,s,n) \delta_{k, r-s-n} \bs_m s_n s_s 
+ \delta_{k, m-n} \delta_{n, r-s} (m+n) (r+s) \bs_m \bs_{s}   s_r
\Bigr] .
\label{A9d}
\eeqar
In (\ref{A9c}) and (\ref{A9d}), $\sigma (r, s, n)$ is given by
\beq
\sigma (r, s, n) = {1\over 2} \left( 2\, r^2 - s^2 - n^2 + r ( n+s) + 2 n s\right) .
\label{A9e}
\eeq
This is symmetric in $n, s$. In ${\rm Term~} 3$, $\chi^{(3)}_k$ can be removed by a suitable choice
of $\bw^{(3)}_k$, making the coefficient of
$L_k$  to be an antiholomorphic form, except for the last term in (\ref{A9b}).
We also note that there are cubic terms arising from the use of $\bw^{(2)}_k$,
as in (\ref{A7}), in ${\rm Term}~2$. For this, we rewrite ${\bar C}^{(2)}_k $ as
\beqar
{\bar C}^{(2)}_k &=& \sum_l D^{(2)}_{k l} d\bs_{l} - d \chi^{(2)}_k +
{\tilde{\bar C}}^{(2)}_k ,\nonumber\\
{\tilde{\bar C}}^{(2)}_k&=& - {1\over 2}\biggl[
\sum \Bigl[ \delta_{k, m+n} \delta_{n, r-s} (m-n) (r+s) \bs_m( d \bs_r s_s + \bs_r ds_s)\Bigr]\nonumber\\
&&+ \sum  (m+n) (r+s) \Bigl[\delta_{k, m-n} \delta_{m, r-s}  s_n (d \bs_r s_s + \bs_r ds_s)
+ \delta_{k, m-n} \delta_{n, r-s}  \bs_r s_s d\bs_m\nonumber\\
&&\hskip .3in - \delta_{k, m-n} \delta_{n, r-s}  \bs_m ( d\bs_r s_s +
\bs_r ds_s )
- \delta_{k, m-n} \delta_{m, r-s}  \bs_r s_s ds_n \Bigr]\biggr] .
\label{A9f}
\eeqar
In this expression, terms of the form $\bs \bs ds$ can be written as
$d [ \bs \bs s] - d\bs \bs s - \bs d\bs s$;
the total derivative adds to the expression for $\chi^{(3)}_k$ and is removed by
choice of $\bw^{(3)}_k$. What is left will be an antiholomorphic form.
This does not work for the two terms in
(\ref{A9f}) which have the $\bs s ds$ combination.
These potentially problematic terms can be simplified as
\beqar
{\rm Problematic~ terms ~in}~{\tilde{\bar C}}^{(2)}_k
&=& - {1\over 2} \sum \delta_{k, r-s-n} (r-s+n) (r +s) \bs_r ( s_n ds_s - s_s ds_n)\nonumber\\
&=&- {1\over 2} \sum \delta_{k, r-s-n} ( n+s+ r+ (n -s) \bs_r s_n ds_s ,
\label{A9g}
\eeqar
where, in the second line, we have used the antisymmetry of the first term in $n, s$
to simplify the result.
Comparing the contribution of the last term in (\ref{A9b}) to $(1/3! ) {\bar C}^{(3)}_k$
and the contribution of (\ref{A9g}) to $- (1/2!) {\bar C}^{(2)}_k$, we see that they cancel out
exactly. After removal of $\chi^{(3)}_k$ terms via choice of
$\bw^{(3)}_k$, we see that what is left of
${\bar C}^{(3)}_k $ is an antiholomorphic form.

The coefficient of $L_k$ can thus be written as
\beq
{\rm Coefficient ~of~} L_k = \bcalE^k = d\bs_\bk - {1\over 2} \sum_l D^{(2)}_{k l} d\bs_l
+ {1\over 3!} \sum_l D^{(3)} d\bs_l ~+ \cdots .
\label{A9h}
\eeq

We have thus verified, in an expansion around the origin to cubic order in the coordinates,
that there is a choice of
$w_n$, $\bw_n$ as a function of the coordinates $s_n$ , $\bs_\bn$
for which $\bcalE^n$ is an antiholomorphic one-form. 
\vskip.05in\noindent
\underline{Maurer-Cartan relations}

The analysis given above for small values of $s_k$, $\bs_k$ shows that one can choose
$\bcalE^n$ to be an anti-holomorphic one-form in an infinitesimal neighborhood
of the origin. Our aim is now to extend this to larger and larger regions
by a sequence of translations, $U \rightarrow U V$ where
$V$ is as in (\ref{10}).  For this, we will also need to use the Maurer-Cartan identity
for $U$, so we will first work this out.
From (\ref{4}), assuming that we have already obtained the holomorphicity
properties for $\E^n$ and $\bcalE^n$, we can write 
\beqar
U^\dagger \del U &=& - \sum_n \E^n L_{-n} + \E^0 \, L_0 + \E {\mathbb 1} , \nonumber\\
U^\dagger \bdel U &=&  \sum_n \bcalE^n L_{n} - \bcalE^0 \, L_0 - \bcalE {\mathbb 1} .
\label{A10}
\eeqar
Taking the holomorphic exterior derivative of the first of these equations, we get
one of the Maurer-Cartan identities as
\beq
\sum\left[ - \del \E^n L_{-n} + \del \E^0 L_0 + \del \E {\mathbb 1}
+ {1\over 2} (m-n) \E^n\wedge \E^m L_{-n-m} - n \E^0 \wedge \E^n L_{-n}\right] = 0 .
\label{A11}
\eeq
This yields three sets of relations corresponding to the coefficients of
$L_{-n}$, $L_0$ and $\mathbb 1$. These are
\beq
 - (\del \E^n + n\, \E^0 \wedge \E^n ) + {1\over 2}  \sum_{r,s} (s-r) \,\delta_{r+s,n}
\, \E^r \wedge \E^s  =  0,
\label{A12}
\eeq
\beq
\del \E^0 = 0, \hskip .3in \del \E = 0 .
\label{A13}
\eeq
The last two relations tell us that we can write
\beq
\E^0 = {1\over 2} \del W^0, \hskip .3in
\E = {1\over 2}  \del W .
\label{A14}
\eeq
Notice that the K\"ahler potential is related to these as
$K = W^0 h + W $. We can now use these expressions for $\E^0$, $\E$ to
write (\ref{A12}) as
\beq
- \del {\tilde \E}^n + {1\over 2} \sum_{r,s} (s-r) \delta_{r+s, n} \, {\tilde\E}^r \wedge 
{\tilde \E}^s = 0 ,
\label{A15}
\eeq
where ${\tilde \E}^n = \E^n \exp(n W^0/2)$. This is the key identity we will need
for extending the previous result.
\vskip.05 in\noindent
\underline{Extending the result by use of translational invariance}

Now consider defining $\E$'s and $\bcalE$'s after translation by $V$.
To first order in 
$\sum {\bar \xi}_n  \, L_n - \xi_n L_{-n}$, this leads to
\beqar
(U V)^{-1} d (U V) &=& V^{-1} dV + V^{-1} ( U^{-1} dU) V\nonumber\\
&\approx& U^{-1} dU + \sum \left[ d{\bar \xi}_n  \, L_n - d \xi_n L_{-n}
- [ {\bar \xi}_n  \, L_n - \xi_n L_{-n}, U^{-1} dU ] \right] .
\label{A16}
\eeqar
We want to show that the coefficient of $L_n$, $n >0$ is an anti-holomorphic
one-form; i.e., one can choose $\bxi_n$ such that
 the holomorphic differential part vanishes.
The condition for this, upon using
(\ref{4}) and evaluating the commutator term, becomes
\beqar
\sum\left[  \del {\bar \xi}_n  \, L_n 
+ \bxi^n \E^m (m+n) L_{n-m}
- \bxi^n \E^0  n L_n \right] = 0,
\label{A17}
\eeqar
with $n>m$. Rewriting this by isolating the coefficient of $L_k$, we get
\beq
 \del {\bar \xi}_k - k\, \E^0 \,\bxi_k
+ \sum (2 m +k )\, \bxi_{m+k} \,  \E^m  = 0 .
\label{A18}
\eeq
Defining ${\tilde \bxi}_k = \bxi_k \exp(- k W^0/2)$, we can further write these conditions
as
\beq
\del\, {\tilde\bxi}_k + \sum_m (2 m +k) {\tilde \bxi}_{m+k} \, {\tilde \E}^m = 0 .
\label{A19}
\eeq
These are to be regarded as a set of equations which can be solved for
$\bxi_k$. However, there are integrability conditions for these equations.
They correspond to taking another holomorphic exterior derivative
of (\ref{A19}), and upon using (\ref{A19}) again, become
\beq
\sum (2 m+k ) \left[
{\tilde \bxi}_{m+k}  \del {\tilde \E}^m - \sum_r ( 2 r + m +k ) 
{\tilde \bxi}_{m+k+r} {\tilde \E}^r\wedge {\tilde \E}^m\right] = 0 .
\label{A20}
\eeq
Because of the wedge product, we can antisymmetrize the coefficient of
${\tilde \E}^r\wedge {\tilde \E}^m$ in $r,m$. For this, we can use
\beq
{1\over 2} \left[ (2m +k)  ( 2 s + m +k) - ( m \leftrightarrow s) \right] 
= {1\over 2} (m-s) \left[ 2 (m+s) + k \right] .
\label{A21}
\eeq
Further we take
$m \rightarrow n$ in the first term and $m \rightarrow s$, $m+r \rightarrow n$ in the second term.
Equation (\ref{A20}) can then be written as
\beq
\sum (2 n +k) {\tilde \bxi}_{n+k} 
\left[ \del {\tilde \E}^n - {1\over 2} \sum_{r,s} (s-r) \delta_{r+s, n} {\tilde \E}^r \wedge {\tilde \E}^s
\right] = 0 .
\label{A22}
\eeq
These are obviously satisfied as a result of the Maurer-Cartan identity
(\ref{A15}).

What we have shown is that if we have $U$ with a choice of $w_n$, $\bw_n$
as functions of $s_k$, $\bs_k$
for which $\E^n$ is a holomorphic one-form and $\bcalE^n$ is an antiholomorphic
one-form, then we can find $\xi_n$, $\bxi_n$ such that $(UV)^{-1} d(UV)$
will have a holomorphic one-form as the coefficient of $L_{-n}$ and an anti-holomorphic
one-form as the coefficient of $L_n$.
This result, combined with the previous result that this property can be obtained
in an infinitesimal neighborhood of the origin, as shown by explicit power series
expansion, shows we can find coordinates such that $\E^n$ is a $(1,0)$-form and
$\bcalE^n$ is a $(0,1)$-form.

 %%%%%%%%%%%%%%%%%%%%%%%%%%%%%%%%%%%%%%%%%%%%%%%
%%%%%%%%%%%%%%%%%%%%%%%%%%%%%%%%%%%%%%%%%%%%%%%
%%%%%%%%%%%%%%%%%%%%%%%%%%%%%%%%%%%%%%%%%%%%%%%
%%%%%%%%%%%%%%%%%%%%%%%%%%%%%%%%%%%%%%%%%%%%%%%
%%%%%%%%%%%%%%%%%%%%%%%%%%%%%%%%%%%%%%%%%%%%%%%
%%%%%%%%%%%%%%%%%%%%%%%%%%%%%%%%%%%%%%%%%%%%%%%

%%%%%%%%%%%%%%%%%%%%%%%%%%%%%%%%%%%%%%%%%%%%%%%
%%%%%%%%%%%%%%%%%%%%%%%%%%%%%%%%%%%%%%%%%%%%%%%

%%%%%%%%%%%%%%%%%%%%%%%%%%%%%%%%%%%%%%%%%%%%%%%
%%%%%%%%%%%%%%%%%%%%%%%%%%%%%%%%%%%%%%%%%%%%%%%
%%%%%%%%%%%%%%%%%%%%%%%%%%%%%%%%%%%%%%%%%%%%%%%
%%%%%%%%%%%%%%%%%%%%%%%%%%%%%%%%%%%%%%%%%%%%%%%
%%%%%%%%%%%%%%%%%%%%%%%%%%%%%%%%%%%%%%%%%%%%%%%
%%%%%%%%%%%%%%%%%%%%%%%%%%%%%%%%%%%%%%%%%%%%%%%
%%%%%%%%%%%%%%%%%%%%%%%%%%%%%%%%%%%%%%%%%%%%%%%
%%%%%%%%%%%%%%%%%%%%%%%%%%%%%%%%%%%%%%%%%%%%%%%
\end{document}